\documentclass[a4paper,fleqn,usenatbib]{mnras}

\usepackage{newtxtext,newtxmath}

\usepackage[T1]{fontenc}
\usepackage{ae,aecompl}

\usepackage{graphicx}	
\usepackage{amsmath}	
\usepackage{amssymb}	
\usepackage{upgreek}
\usepackage{aas_macros} 
\usepackage{url}
\usepackage{booktabs}

\title[r~modes in upper main-sequence stars]{Theory and evidence of global Rossby waves in upper 
 main-sequence stars: r-mode oscillations in many {\it Kepler} stars}

\author[H. Saio et al.]{
 Hideyuki Saio$^{1}$\thanks{E-mail: saio@astr.tohoku.ac.jp (HS)},
Donald W. Kurtz$^{2}$,
Simon Murphy$^{3,4}$,
Victoria L. Antoci$^{4}$,
\newauthor{
and Umin Lee$^{1}$
}
\\
$^{1}$Astronomical Institute, Graduate School of Science, Tohoku University, Sendai, Miyagi 980-8578, Japan\\
$^{2}$Jeremiah Horrocks Institute, University of Central
Lancashire, Preston PR1 2HE, UK\\
$^{3}$Sydney Institute for Astronomy, School of Physics, The University of Sydney, NSW 2006, Australia\\
$^{4}$Stellar Astrophysics Centre, Department of Physics and Astronomy, Aarhus University, Ny Munkegade 120, DK-8000 Aarhus C, Denmark\\
}

\date{Accepted XXX. Received YYY; in original form ZZZ}

\pubyear{2017}

\begin{document}
\label{firstpage}
\pagerange{\pageref{firstpage}--\pageref{lastpage}}
\maketitle

\begin{abstract}
Asteroseismic inference from pressure modes (p~modes) and buoyancy, or gravity, modes (g~modes) is ubiquitous for stars across the Hertzsprung--Russell diagram. Until now, however, discussion of r~modes (global Rossby waves) has been rare. Here we derive the expected frequency ranges of r~modes in the observational frame by considering the visibility of these modes.  We find that the frequencies of r~modes of azimuthal order $m$ appear as groups at slightly lower frequency than $m$ times the rotation frequency. Comparing the visibility curves for r~modes with Fourier amplitude spectra of {\it Kepler} light curves of upper main-sequence B, A and F stars, we find that r~modes are present in many $\gamma$~Dor stars (as first discovered by \citealt{vanr16}), spotted stars, and so-called Heartbeat stars, which are highly eccentric binary stars. We also find a signature of r~modes in a frequently bursting Be star observed by  {\it Kepler}.  In the amplitude spectra of moderately to rapidly rotating $\gamma$~Dor stars, r-mode frequency groups appear at lower frequency than prograde g-mode frequency groups, while in the amplitude spectra of spotted early A to B stars, groups of symmetric (with respect to the equator) r-mode frequencies appear just below the frequency of a structured peak that we suggest represents an approximate stellar rotation rate. In many Heartbeat stars, a group of frequencies can be fitted with symmetric $m=1$ r~modes,  which can be used to obtain rotation frequencies of these stars.  
\end{abstract}




\section{Introduction}

Global normal modes of Rossby waves in rotating stars were first named r~modes by \citet{pap78}. 
These modes consist of predominantly toroidal motions that cause no restoring force nor light variations in a non-rotating star because they cause no compression or expansion.
However, in a rotating star, the toroidal motion couples with spheroidal motion caused by the Coriolis force, and in that case r~modes cause temperature perturbations, hence are visible.

A simple explanation for the generation of Rossby waves may be given by considering the conservation of the vertical component of the total vorticity $[(\nabla\times\bmath{u})_r+\Omega_r]$ \citep{ped82}, where $\bmath{u}$ is the velocity in the co-rotating frame and $\Omega_r=\Omega\cos\theta$ is the radial component of the angular velocity of rotation with $\theta$ being co-latitude. With the convention that the rotation axis points to the north, if we follow the northward motion of a fluid element in northern hemisphere (say at point A), $\Omega_r$ increases, and due to conservation of the vertical vorticity a clockwise motion is generated around A, which causes a northward motion of a fluid element (say B) located to the west of A.  Then, for the same reason as before, a clockwise motion is generated around B, which pushes back the fluid element A to the original position and at the same time moves the west-side fluid element B northward. In this way, waves propagate westward, i.e. in the direction retrograde to rotation \citep[see also][]{sai82}.  For this reason, r~modes are always retrograde in the co-rotating frame. Moreover, because the gradient of $\Omega_r$ is larger in mid-latitudes than at the equator, r~modes tend to be confined to the mid-latitude  range on the surface \citep{lee97}.

Because of coupling with spheroidal motions that cause density -- and hence pressure -- perturbations, r~modes have an infinite number of discrete (low) frequencies and corresponding eigenfunctions, just as g~modes do. Adiabatic frequencies and eigenfunctions for r~modes in a stellar model were first obtained by \citet{pro81}. This coupled spheroidal motion also makes $\kappa$-mechanism excitation of r~modes possible, as shown theoretically for DA white dwarfs with excitation in the H-ionisation zone \citep{sai82,ber83} and for Slowly Pulsating B (SPB) stars with excitation in the Fe opacity bump zone \citep{tow05,sav05,lee06}. Searches for  observational evidence of r~modes in stars had been unsuccessful \citep[e.g.,][]{rob82,kep84} until the discovery by \citet{vanr16}, who identified r~modes in many moderately to rapidly rotating  $\gamma$~Dor stars from the period-spacing and period relations. 

In other stars, r~modes are generally believed to be present in neutron stars -- because of the instability caused by gravitational wave radiation -- and to play an important role in spinning down neutron stars 
\citep[e.g.,][]{and98,bro00,mah13}, but they have not been observed. Even the slowly rotating Sun seems to have r~modes \citep[e.g.,][]{stu10,stu15}. 

In this paper, we argue that r~modes are common in spotted stars, stars in binaries, and g-mode pulsators for B, A and F stars on the upper main-sequence. We show a close agreement in frequency range between our theoretical calculations and humps of unresolved peaks in the amplitude spectra of the ultra-precise light curves of many stars observed by the {\it Kepler} space telescope.

\section{Basic properties of r~modes}
\subsection{Slow rotation limit}

In the case of very slow rotation ($\Omega \ll \sqrt{GM/R^3}$), the displacement of r~mode oscillations in the co-rotating frame can be written as
\begin{equation}
\bmath{\xi} \simeq e^{i\sigma t}\left[ T^m_\ell(r)\left(0,{1\over\sin\theta}{\partial\over\partial\phi}, -{\partial\over\partial\theta}\right)Y_\ell^m(\theta,\phi) + \bmath{S}^m \right] ,
\end{equation}
where $G$ is the gravitational constant, $M$ is stellar mass, $R$ is radius,  $\sigma$ is the angular frequency of pulsation in the co-rotating frame, and $Y_\ell^m(\theta,\phi)$ is a spherical harmonic, which is proportional to $\exp(im\phi)$. In this paper, we adopt the convention that the frequency is always positive (i.e., $\sigma > 0$) so that a positive $m \  ( > 0)$ represents a {\it retrograde} mode and a negative $m$ a prograde mode.  The term proportional to $T^m_\ell(r)$ represents (dominant) toroidal displacements \citep[some examples of flow patterns on the surface are presented in][]{sai82}, while $\bmath{S}^m$ stands for associated small spheroidal displacements of the  order $\Omega^2$ \citep{pro81}. The spheroidal term $\bmath{S}^m$ may be represented as
\begin{equation}
\bmath{S}^m = \sum_{\ell_{\rm s} } \left(R^m_{\ell_{\rm s}}(r), ~H_{\ell_{\rm s}}^m(r){\partial\over\partial\theta}, ~{H_{\ell_{\rm s}}^m(r)\over\sin\theta}{\partial\over\partial\phi}\right)Y^m_{\ell_{\rm s}}(\theta,\phi),
\end{equation}  
where $R^m_{\ell_{\rm s}}(r)$ and $H_{\ell_{\rm s}}^m(r)$ are the amplitudes of radial and horizontal spheroidal displacement.
The summation means a sum of terms with $\ell_{\rm s}=\ell -1$ and $\ell+1$, although $\ell-1$ terms diminish if $\ell = m$. This rule comes from the fact that toroidal motions associated with $Y_\ell^m$ couple, due to the Coriolis force, with the  spheroidal motions associated with $Y_{\ell +1}^m$ and $Y_{\ell-1}^m$ (the latter is non-zero only if $\ell > m$).\footnote{If we consider higher order terms, the generated spheroidal motions of $\ell_{\rm s} \ (=\ell\pm1$) couples, in turn, with toroidal motions associated with 
$\ell_{\rm s} \pm 1 ~  (= \ell-2, \ell, {\rm and}~ \ell+2)$, and so on. Therefore, if higher order terms are included, an odd r~mode involves toroidal terms of $\ell = m, m+2, m+4, \ldots$ and spheroidal terms of $\ell_s = m+1, m+3, m+4, \ldots$, while an even r mode involves toroidal terms of $\ell = m+1, m+3, m+5, \ldots$ and spheroidal terms of $\ell_s = m, m+2, m+4, \ldots$ \citep{lee95}.}
 
Thus, the latitudinal degree of the spheroidal term, $\ell_{\rm s}$, is different by $\pm 1$ from $\ell$. Since, the spheroidal terms produce temperature perturbations, we designate even and odd modes with respect to $\ell_{\rm s}$; i.e., if $(\ell_{\rm s}-m)$ is zero or an even number [i.e., $(\ell - m)$ is a odd number] we call the mode an even mode, which produces temperature perturbations symmetric to the equator (i.e., the temperature perturbations in the northern and southern hemispheres are in phase with each other), and vice versa. 

In the slow rotation limit of $\Omega \ll \sqrt{GM/R^3}$, the frequency of an r~mode in the co-rotating frame, $\nu^{\rm co}_{\rm rm}$, can be written as
\begin{equation}
\nu^{\rm co}_{\rm rm} \approx \nu_{\rm rot}\left[{2m\over \ell(\ell+1)} + q(n){\Omega^2\over GM/R^3}\right] 
\end{equation}
\citep{pap78}, where $\nu_{\rm rot} =\Omega/(2\pi)$ and $q(n)$ is a non-dimensional quantity. 
Since $\ell \ge m \ge 1$ for r~modes, we have the inequalities 
\begin{equation}
{2m\nu_{\rm rot}\over \ell(\ell+1)} \le {2\nu_{\rm rot}\over m+1} \le \nu_{\rm rot}.
\end{equation}
The maximum value occurs at $m=1$, for which $q(n)$ is negative and its absolute value increases with the radial order $n$ \citep{sai82}.
Therefore, we have an inequality for the frequencies of r~modes in the co-rotating frame as
\begin{equation}
\nu^{\rm co}_{\rm rm} < \nu_{\rm rot}.
\label{eq:nu_co}
\end{equation}
This relation is important in this paper for our interpretation of the amplitude spectra of the light curves of {\it Kepler} stars.

\subsection{General case under the Traditional Approximation of Rotation}

If rotation is fast in the sense that $\Omega$ is a few tenths of $\sqrt{GM/R^3}$, and the spin parameter, $2\nu_{\rm rot}/\nu^{\rm co}$, is larger than one,  
 we can no longer use $\ell$ and $\ell_{\rm s}$ to describe the properties of low frequency oscillations, because terms proportional to $Y_l^m(\theta,\phi)$ (where $l = \ell$ or $\ell_{\rm s}$) couple significantly with terms proportional to $Y_{l\pm2}^m(\theta,\phi)$. To discuss the properties of such low-frequency oscillations in a moderately or rapidly rotating star,  the Traditional Approximation of Rotation (TAR) is useful. In this approximation, we assume uniform rotation, adopt the co-rotating frame, and neglect the horizontal component of angular velocity of rotation, $\Omega\sin\theta$ with $\theta$ being co-latitude. In addition, we neglect centrifugal deformation (which is not very important for low-frequency oscillations, whose propagation cavity resides in the deep interior), and use the Cowling approximation, in which the Eulerian perturbation of the gravitational potential is neglected. Neglecting $\Omega\sin\theta$ corresponds to neglecting the Coriolis force associated with radial motions, and neglecting the radial component of Coriolis force associated with horizontal motions. The TAR is a good approximation for low-frequency pulsations in which horizontal motions dominate. 

\begin{figure}
\includegraphics[width=\columnwidth]{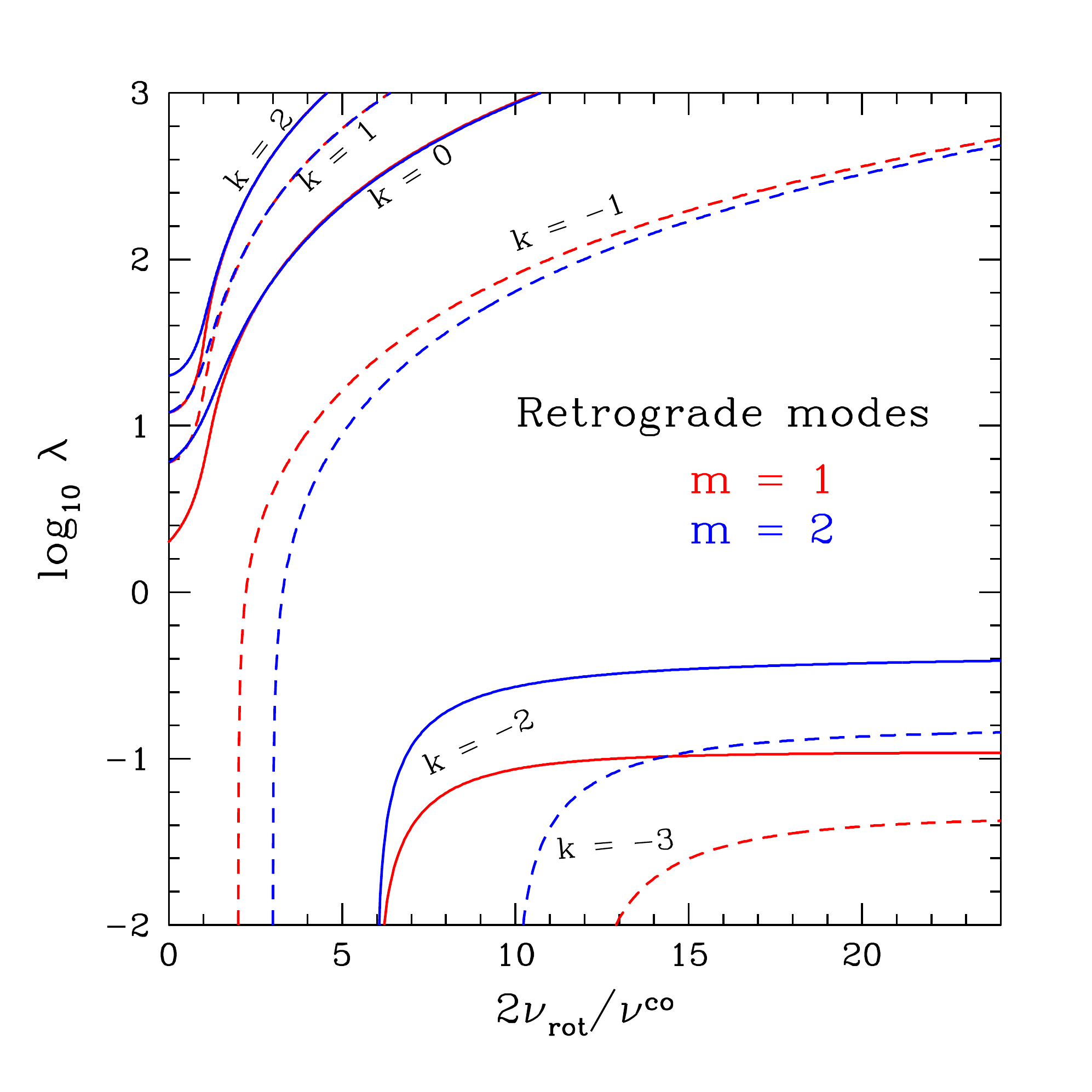}
\caption{The eigenvalue $\lambda$ of Laplace's tidal equation for retrograde modes versus spin parameter $2\nu_{\rm rot}/\nu^{\rm co}$, where $\nu_{\rm rot}$ and $\nu^{\rm co}$ are rotation frequency and pulsation frequency in the co-rotating frame, respectively. The integer $k$ represents the type of mode; $k \le -1$ are for r~modes, and $k \ge 0$ are for g~modes, for which $\lambda = (|m|+k)(|m|+k+1)$, that is, $\ell_{\rm s}=|m|+k$, at $\nu_{\rm rot} = 0.0$. For r~modes $\lambda > 0$ if $2m\nu_{\rm rot}/\nu^{\rm co} > (m + |k|-1)(m+|k|)$; furthermore, $\lambda$ approaches a constant value of  $m^2(2|k|-1)^{-2}$ if $k \le -2$ and $2\nu_{\rm rot}/\nu^{\rm co} \gg 1$ \citep{tow03a}. Even and odd modes are represented by solid and dashed lines, respectively. }
\label{fig:lambda}
\end{figure}

The angular dependence of oscillation, which depends on the spin parameter, is obtained by solving Laplace's Tidal equation \citep[e.g.,][]{lee97} with an eigenvalue $\lambda$. For g~modes, $\lambda \rightarrow \ell_{\rm s}(\ell_{\rm s}+1)$ as $2\nu_{\rm rot}/\nu^{\rm co} \rightarrow 0$, while for r~modes, $\lambda \rightarrow 0$ as $2\nu_{\rm rot}/\nu^{\rm co} \rightarrow \ell(\ell+1)/m$. Diminishing $\lambda$ at a limiting frequency means that the oscillation motion is purely toroidal without any horizontal compression at the limiting frequency. Only if $\lambda > 0$ can r~modes exist\footnote{Since $\ell(\ell+1)/|m| > 2$ in general, the spin parameter of an r~mode is always larger than 2; i.e. $2\nu_{\rm rot}/\nu^{\rm co}_{\rm r\,mode} > 2$.};
 larger radial order corresponds to a larger value of $\lambda$ (and hence a lower frequency in the co-rotating frame) indicating the stronger effects of spheroidal motions (or buoyancy). Fig.\,\ref{fig:lambda} shows $\lambda$ for retrograde modes (we use the convention $m>0$ for retrograde modes in this paper) as a function of spin parameter $2\nu_{\rm rot}/\nu^{\rm co}$ \citep[the prograde part is shown in, e.g.,][]{sai17,lee13,ack10}. As shown in this figure, the type of latitudinal dependence of the oscillations is identified by integers, $m$ and $k$, where ordering by $k$ is adopted from \citet{lee97}. Even and odd symmetries with respect to the equator for the distribution of temperature and pressure variations are designated by $|k|$ -- even if $|k|$ is zero or an even number, while odd if $|k|$ is an odd number. For r~modes $k\le -1$, while for g  modes $k \ge 0$ ($k=\ell_{\rm s} - |m|$ if $\nu_{\rm rot}=0$; $k=0$ for sectoral g~modes). 

\begin{figure}
\includegraphics[width=0.49\columnwidth]{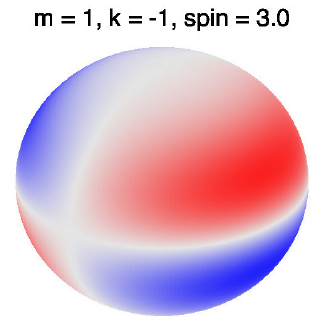}
\includegraphics[width=0.49\columnwidth]{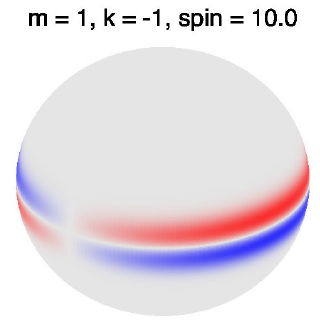}

\vspace{0.5cm}
\includegraphics[width=0.49\columnwidth]{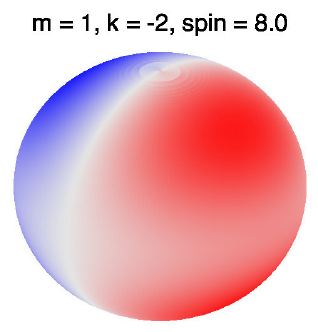}
\includegraphics[width=0.49\columnwidth]{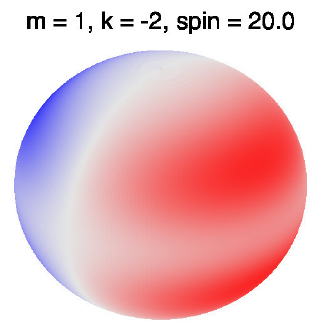}

\vspace{0.5cm}
\includegraphics[width=0.49\columnwidth]{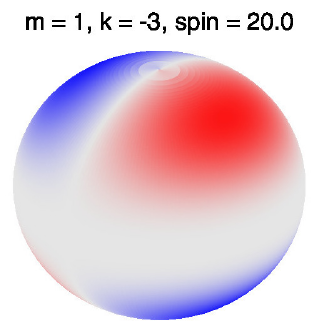}
\includegraphics[width=0.49\columnwidth]{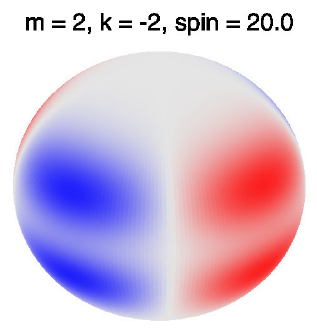}
\caption{Distributions of temperature variations for some r~modes on the stellar surface. Parameters for each mode are given above each pattern, where `spin' means the spin parameter defined as $2\nu_{\rm rot}/\nu^{\rm co}$. Positive, negative, and zero variations are shown by blue, red, and gray, respectively. The inclination angle to the rotation axis is set to be $60^\circ$ for all cases.  }
\label{fig:pattern}
\end{figure}

Fig.\,\ref{fig:pattern} shows distributions of temperature variations on the stellar surface for some r modes with various parameters. Generally, a large amplitude part is concentrated towards a smaller region if the mode has a larger $\lambda$.
In particular, for the odd r modes with $(m,k)=(1,-1)$, the amplitude is strongly concentrated towards the equatorial zone if the spin parameter is much larger than 2; i.e, if the value of $\lambda$ is very large ($\lambda\sim\!\!10^2$ at $2\nu_{\rm rot}/\nu^{\rm co}=10$; see Fig.\,\ref{fig:lambda}).
On the other hand, for the case of even r modes with $(m,k)=(1,-2)$, the amplitude is only slightly concentrated towards mid-latitudes, even if the spin parameter becomes large, because $\lambda$ stays nearly constant.

Under the TAR, the set of equations for non-radial adiabatic pulsations of a rotating star with the Cowling approximation is the same as that for the non-rotating case, except that $\ell_{\rm s}(\ell_{\rm s}+1)$ is replaced with $\lambda$ \citep{unno,lee97,ack10}.\footnote{This is just formally so, because $\lambda$ varies as a function of frequency for a given rotation rate. } This simplifies significantly the calculations of low-frequency nonradial pulsations of a rotating star.
All adiabatic analyses for nonradial pulsations in this paper were performed using the TAR with the standard non-dimensional variables, $y_1=\xi_r/r$ and $y_2=p'/(g\rho r)$ \citep[e.g.,][]{unno}, where $r$, $p', \rho$ and $g$ are the distance from the centre of the star, Eulerian perturbation of pressure, the gas density, and local gravity, respectively. 
 
Furthermore, under the TAR, we can use asymptotic formulae of high order g~modes in non-rotating stars for both r and g~modes in rotating stars,  if $\ell_{\rm s}(\ell_{\rm s}+1)$ is replaced with $\lambda$. Thus, the frequency in the co-rotating frame of a high radial-order r or g~mode in a rotating star can be represented formally as
\begin{equation}
    \nu^{\rm co} \approx {\sqrt{\lambda} \over 2\pi^2n}\int {N\over r} dr  
    \equiv {\sqrt{\lambda}\over n}\nu_0,
\label{eq:freqco}
\end{equation}
\citep[e.g.][]{lee87b,bou13}
where $N$ is the Brunt-V\"ais\"al\"a frequency. The corresponding frequency in the inertial (observer's) frame, $\nu^{\rm int}$ is given as
\begin{equation}
    \nu^{\rm int} = \left|\nu^{\rm co}-m\nu_{\rm rot}\right|
    \approx \left|{\sqrt{\lambda}\over n}\nu_0 - m\nu_{\rm rot}\right|.
\label{eq:freqin}
\end{equation}

\subsubsection{Period spacings}
\label{section:periodspacings}

From equation (\ref{eq:freqin}), taking into account the fact that r-mode frequency in the co-rotating frame is always smaller than $\nu_{\rm rot}$ (eq.\ref{eq:nu_co}), the frequency of an r~mode in the inertial frame, $\nu^{\rm int}_{\rm rm}$ is given as
\begin{equation}
\nu^{\rm int}_{\rm rm} \approx m\nu_{\rm rot} - {\sqrt{\lambda_{\rm rm}}\over n}\nu_0;   \quad \mbox{r~modes} ~ m \ge 1,
\label{eq:rm_inert}
\end{equation}
where the symbol $\lambda_{\rm rm}$ is to emphasize $\lambda$ of r~modes. Note that in the inertial frame a lower radial order $n$ corresponds to a lower frequency (i.e., a longer period), while in the co-rotating frame the relation is the opposite. In other words, as $n \rightarrow \infty$, $\nu_{\rm rm}^{\rm int} \rightarrow m\nu_{\rm rot}$ and $\nu_{\rm rm}^{\rm co} \rightarrow 0$.  

Using equation\,(\ref{eq:rm_inert}) and assuming $\lambda_{\rm rm}$ to be constant (i.e., far from the zero point), we obtain the period spacing of r~modes, $\Delta P^{\rm int}_{\rm rm}$, as
\begin{equation}
\Delta P^{\rm int}_{\rm rm} \approx \left(P^{\rm int}_{\rm rm}\right)^2{\sqrt{\lambda_{\rm rm}}\over n^2}\nu_0 ={\sqrt{\lambda_{\rm rm}}\over (\nu_{\rm rm}^{\rm int}n)^2}\nu_0 ,
\end{equation}
where $n \gg 1$ is assumed. 
If $2\nu_{\rm rot}/\nu^{\rm co}\gg1$, $\lambda_{\rm rm} \approx m^2(2|k|-1)^{-2}$ \citep{tow03a} for $k\le -2$.
This equation indicates that the period spacing of r~modes, $\Delta P^{\rm int}_{\rm rm}$, increases with period in the observer's frame (i.e., decreases with frequency). Note that this relation is derived assuming $\lambda_{\rm rm}$ to be nearly constant, which is not applicable near the limiting frequency where $\lambda$ goes to zero (Fig.~\ref{fig:lambda}), in which case the period spacing diminishes (as $\nu^{\rm co}$ approaches $2m\nu_{\rm rot}/[(m+|k|-1)(m+|k|)]$).

For prograde sectoral g~modes (i.e., $k=0, m<0$) $\lambda$ approaches $m^2$ \citep[e.g.,][]{sai17} as the spin parameter  $2\nu_{\rm rot}/\nu^{\rm co}_g$ increases (i.e., for high order~modes). Assuming $\lambda \simeq m^2$, we can derive the period spacing for high-order, prograde sectoral g~modes similarly as
\begin{equation}
\Delta P^{\rm int}_g \simeq {1\over|m|}{\nu_0\over (\nu_0+n\nu_{\rm rot})^2 }\, ,\qquad
\mbox{for prograde g~modes of}~ k=0.
\end{equation} 
Since g-mode periods increase with radial order $n$, the period spacing of prograde sectoral g~modes decreases with period in the observer's frame. 

\begin{figure}
  \includegraphics[width=\columnwidth]{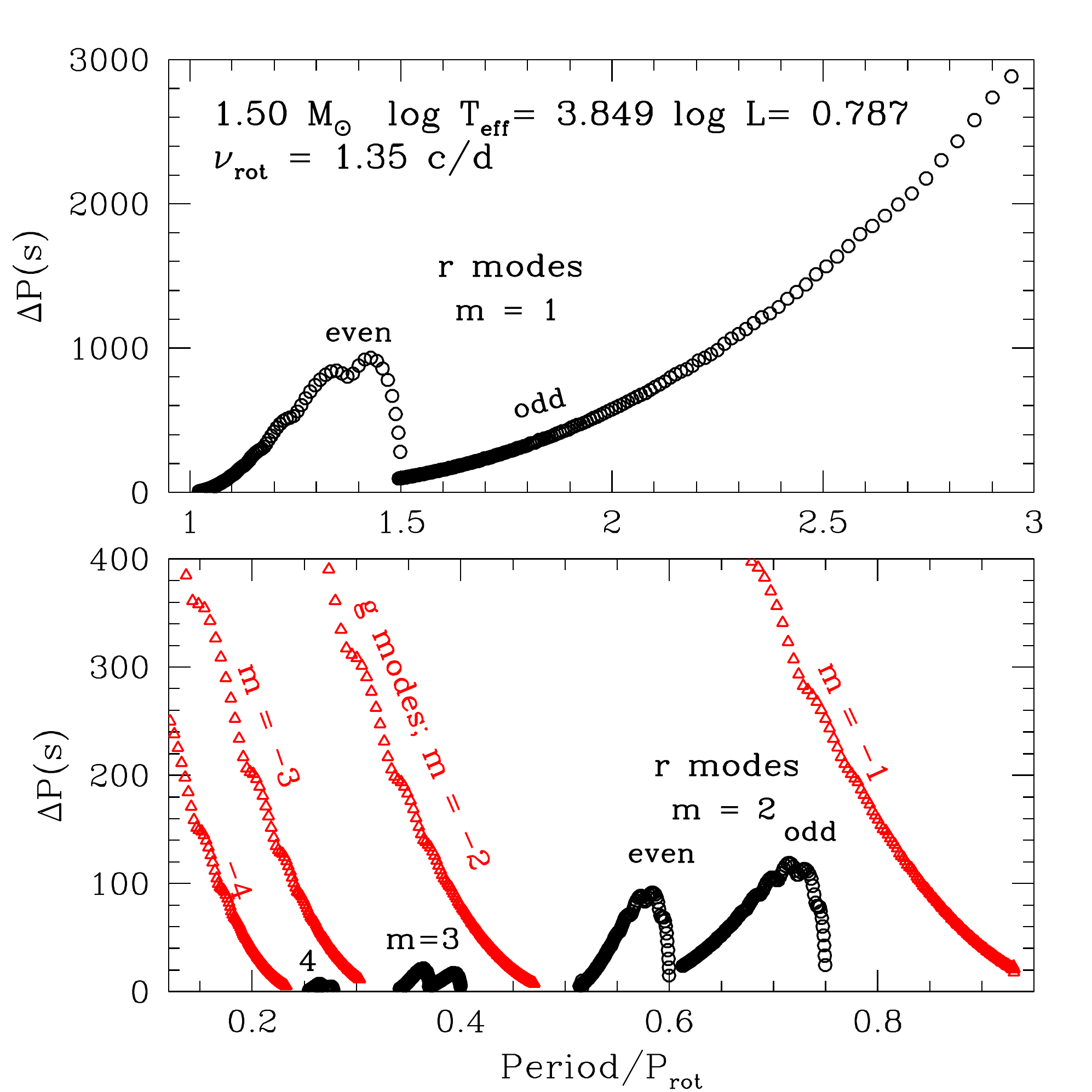}
\caption{Period spacings (in seconds) versus period in the inertial frame (normalized by the rotation period) calculated using the TAR for a 1.5-M$_\odot$  model with a rotation frequency of $1.35$~d$^{-1}$, where the periods and period-spacings are defined in the observer's frame. The top and bottom panels are for the period ranges longer and shorter than the rotation period, respectively. For r~modes, even modes of $k=-2$ and odd modes of $k=-1$ are plotted with open circles, while for g~modes, prograde sectoral modes of $k=0$ are plotted with open triangles. Wiggles in $\Delta P$ are caused by the steep mean molecular weight gradient exterior to the convective core \citep{mig08,bou13}.  } 
\label{fig:dp}
\end{figure}

Fig.\,\ref{fig:dp} shows period spacings for r (open circles) and g (open triangles) modes of a model of 1.5\,M$_\odot$ calculated  using the TAR for a rotation frequency of $1.35$~d$^{-1}$ (chosen as a model of KIC~11907454 discussed below). The rotation speed is about in the middle of the range of those of {\it Kepler} $\gamma$~Dor stars for which \citet{vanr16} detected r~modes. As expected, period spacings of r~modes generally increase with period in the observer's frame, while period spacings of g~modes decrease with period. A rapid decrease in the period spacings of r~modes occurs at the period limit of each r~mode sequence. This is caused by a rapid decrease in $\lambda$ around the limiting period (see Fig.\,\ref{fig:lambda}).

\section{Visibility of r~modes}

We discussed in the previous section that \citet{vanr16} detected many cases of r~modes in $\gamma$\,Dor variables by period spacings alone. However, measuring period spacings is sometimes difficult because of insufficient frequency resolution in the amplitude spectrum  -- even for the 4-year {\it Kepler} data set -- because frequency groups of r~modes often consist of very high-order~modes. In such cases, it is useful to estimate r-mode frequency ranges expected by calculating visibilities, which we discuss in this section.

Under the TAR, the latitudinal amplitude distribution of the temperature and pressure perturbations is described by the solution of Laplace's tidal equation. We write the function as $\Theta^m_k(\cos\theta,\alpha)$, where $\alpha = 2\nu_{\rm rot}/\nu^{\rm co}$ is the spin parameter. The visibility is proportional to the integral of $\Theta^m_k(\cos\theta,\alpha)$ over the projected visible hemisphere of the star. We define the visibility of a mode as
\begin{equation}
\begin{array}{ll}\displaystyle
{\rm Vis.}=y_2\int_0^{2\pi}\!\!\!\! d\phi_{\rm L}\int_0^{\pi/2}\!\! d\theta_{\rm L}\sin(2\theta_{\rm L})[1- \mu(1-\cos\theta_{\rm L})]
\cr \displaystyle \hspace{0.5\columnwidth}
\times~\Theta^m_k(\cos\theta,\alpha)\cos(m\phi),
\end{array}
\label{eq:vis}
\end{equation}
where $y_2$ means the surface value of a non-dimensional pulsational variable proportional to the Eulerian perturbation of pressure, 
$(\theta_{\rm L},\phi_{\rm L})$ are spherical angles defined such that the direction of the line-of-sight is $\theta_{\rm L}=0$, $\phi_{\rm L}$ is the azimuthal angle around the axis, and $\mu$ is the limb-darkening coefficient. We assumed $\mu = 0.6$ in the calculations presented in this paper.
The spherical angles $(\theta,\phi)$ associated with the stellar rotation axis are converted to $(\theta_{\rm L},\phi_{\rm L})$ by spherical trigonometry \citep[e.g.][]{sma77}.

The kinetic energy of an oscillation mode is calculated as
\begin{equation}
{\rm K.E.} = {\sigma^2\over 2} \int_V\bmath{\xi}^*\cdot\bmath{\xi}\rho dV,
\label{eq:ke}
\end{equation} 
where $\bmath{\xi}$ is the displacement vector; the form of each component is given in \citet{lee97}.

\begin{figure}
  \includegraphics[width=\columnwidth]{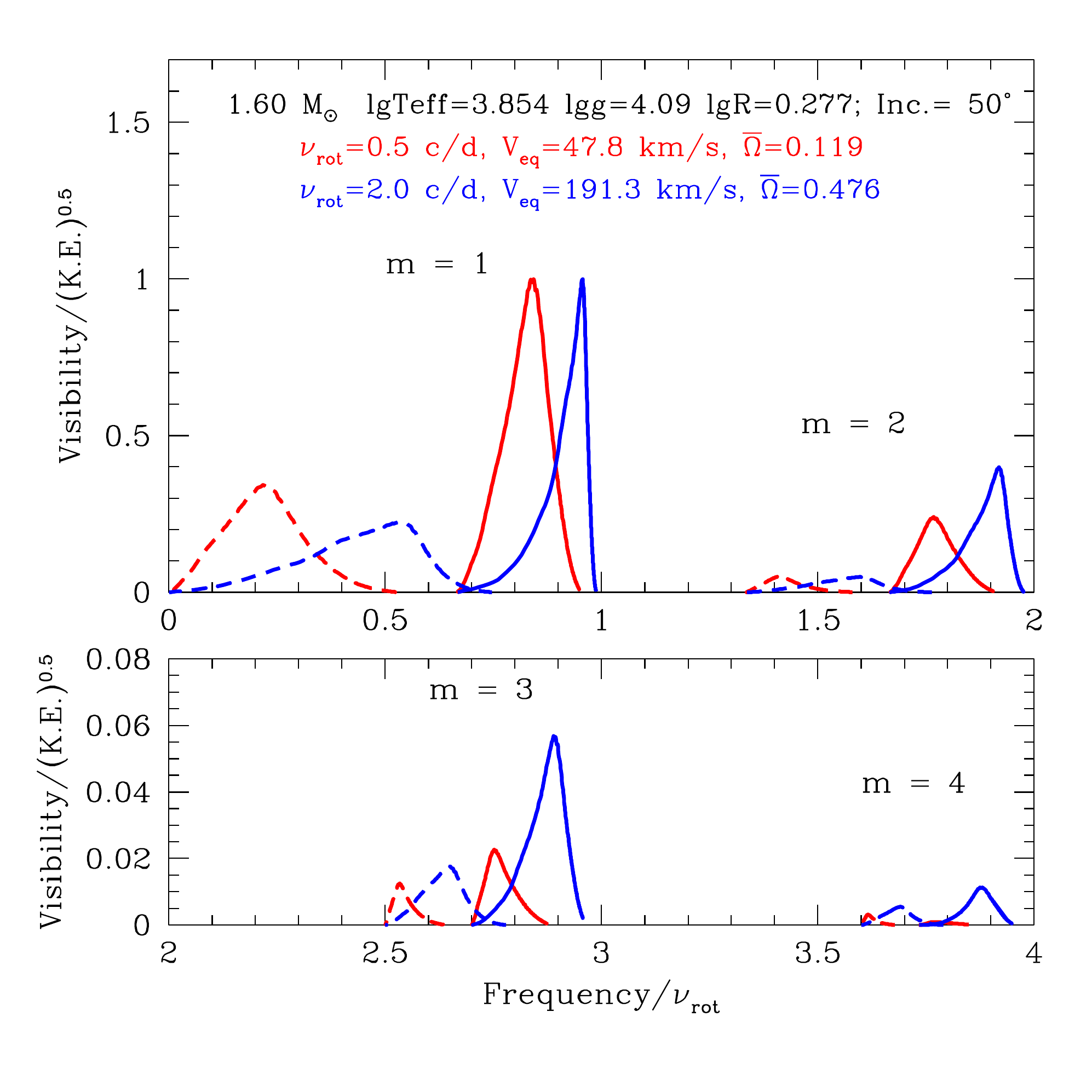}
\caption{Visibilities divided by the square root of kinetic energy of r~modes for $m=1, 2$ (top panel) and $m=3,4$ (bottom panel; note the scale change) are plotted as a function of the frequency in the inertial frame divided by the rotation frequency for two cases of assumed rotation frequencies of $0.5$\,d$^{-1}$ (red lines) and $2.0$~d$^{-1}$ (blue lines); the corresponding equatorial velocities are $47.8$~km~s$^{-1}$ and $191$~km~s$^{-1}$, respectively.
Corresponding angular frequency of rotation normalized as  $\overline{\Omega}\equiv\Omega/\sqrt{GM/R^3}$ is also indicated (critical rotation corresponds to $\overline{\Omega} \approx 0.7$).  
Solid and dashed lines are for even ($k=-2$) and odd  ($k=-1$) modes. For each rotation frequency the maximum value is normalized to unity. An inclination (between rotation axis and line of sight) of $50^\circ$ is adopted as a typical value. 
}
\label{fig:vis_comp}
\end{figure}

The visibility defined by equation(\ref{eq:vis}) is proportional to the amplitude of oscillation, which cannot be determined by linear analysis. To normalize the effect of unknown amplitude, we divide the visibility by the square root of kinetic energy (K.E., which is proportional to the square of amplitude); i.e., we consider Vis./$\sqrt{{\rm K.E.}}$ to represent the observational relative amplitude distribution among different modes in a star if energy is equally distributed among the modes. Fig.\,\ref{fig:vis_comp} compares Vis./$\sqrt{{\rm K.E.}}$ for r~modes of different $(m,k)$ in a 1.6-M$_\odot$ model at two different rotation frequencies; pulsation frequencies in the inertial frame are normalized by the rotation frequency in the horizontal axis.

The amplitude peak occurs around the middle of each sequence at intermediate to high radial orders ($\sim\!\!50 - 100$) depending on the models (higher orders tend to occur in models with a higher mean density). With increasing rotation speed, the frequency at the peak for each sequence of even $m$ modes gets closer to $m\nu_{\rm rot}$ and the amplitude distribution around the peak becomes strongly asymmetric with a steep high-frequency side. The amplitude hump of odd ($k=-1$) modes (dashed line)  is always located at the lower frequency side of the even $k=-2$ modes for a given $m$, and the peak frequency increases with rotation. 
Even r~modes of  $(m,k)=(1,-2)$  (the most visible case), in a moderately to rapidly rotating star are expected in a frequency range of  $0.8\la\nu^{\rm int}/\nu_{\rm rot}\la 1$ (Fig.\,\ref{fig:vis_comp}), in which the lower bound increases with rotation speed. This range corresponds to $\nu^{\rm co} \la 0.2\nu_{\rm rot}$ (eq.\,\ref{eq:freqin}) and hence $2\nu_{\rm rot}/\nu^{\rm co} \ga 10$. The spin-parameter range is consistent with those for r modes found in $\gamma$ Dor stars \citep{vanr16,aer17}. 

As expected, amplitude peaks decrease steeply with increasing $m$; this tendency is stronger in the case of slower rotation. For this reason, detecting r~modes of $m \ge 3$ would be very difficult in most cases -- in particular, in slow rotators -- unless considerable excess energy is given to modes of high $m$. The frequency difference between $k=-1$ (odd, dashed lines) and $k=-2$ (even, solid lines) is larger in the slower rotating case.  We note that the visibility of other odd modes with $(m,k)=(1,-3)$ is much smaller than that of even r~modes with $(m,k)=(1,-2)$, 
although the frequencies in the observational frame are similar. For this reason we do not consider $k\le -3$ modes in this paper. 

\section{Examples from Kepler data}

In this section we compare our theoretical predictions discussed in the previous sections with observational properties derived from {\it Kepler} photometric data. Stellar structure models were calculated without including rotation using the \textsc{MESA} (Modules for Experiments in Stellar Astrophysics; version 4298, 7184) code \citep{pax13}.  Adiabatic pulsation frequencies including the effect of rotation with the TAR were obtained using a pulsation code modified from the code discussed in \citet{sai80}, in which $l(l+1)$ is replaced with $\lambda$.  In the iterative process of obtaining a frequency, the value of $\lambda$ is renewed by interpolating a table that was calculated separately as a function of spin parameter. 

{\it Kepler} long cadence light curve ascii data of each star were downloaded from KASOC (Kepler Asteroseismic Science Operations Center)  website (http://kasoc.phys.au.dk/index.php). Data for each quarter were divided by the mean value and converted  to magnitude differences. Combined data were analysed using the software \textsc{PERIOD04} \citep{period04}. 

\subsection{$\gamma$~Dor stars}
\label{sec:gammaDor} 

\subsubsection{KIC~11907454}

\begin{figure}
  \includegraphics[width=\columnwidth]{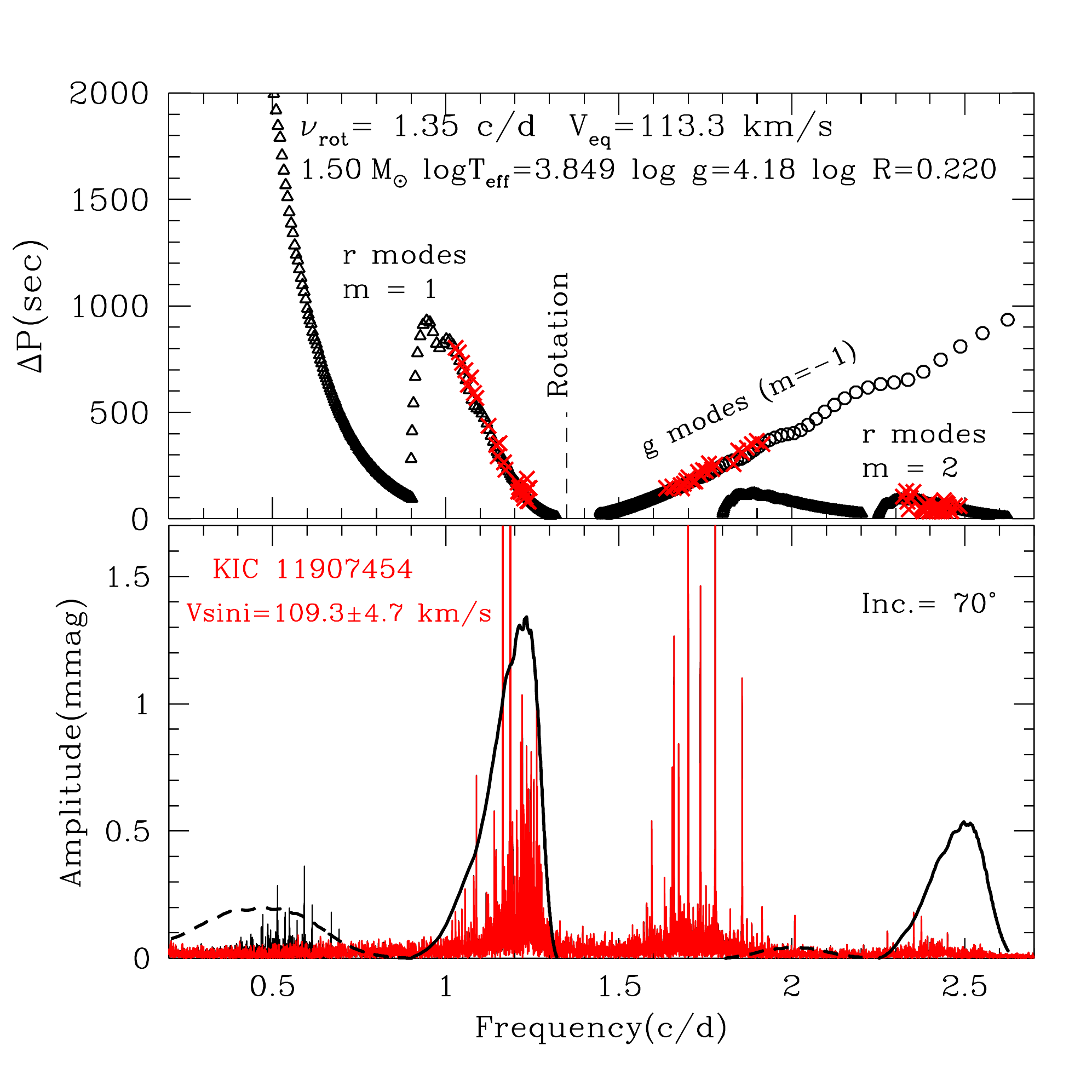}
\caption{Observed period spacings (crosses; top panel) and an amplitude spectrum for the $\gamma$~Dor star KIC~11907454 (bottom panel) are compared with predicted period spacings of r (triangles) and g~modes (circles; top panel) and the expected amplitude distribution of r~modes for an inclination angle of $70^\circ$ (bottom panel). Black solid  lines (for even modes) and dashed lines (for odd modes) show predicted amplitude distributions normalized arbitrarily. The adopted rotation frequency is indicated by a vertical dashed line (top panel), which agrees with  the rotation frequency $1.35\pm0.02$~d$^{-1}$ obtained for KIC~11907454 by \citet{vanr16}. The black part of the amplitude spectrum (bottom panel) indicates beat frequencies (i.e., $|\nu_i-\nu_j|$) between r~modes of $(m,k)=(1,-2)$ and g~modes of $m=-1$. }
\label{fig:gamDor}
\end{figure}

Fig.\,\ref{fig:gamDor} compares the predicted period spacings of r and g~modes and amplitude distributions of r~modes with those of the $\gamma$~Dor star KIC~11907454. The theoretical period spacings are for the same model as in Fig.~\ref{fig:dp}, but are plotted as a function of frequency (rather than period) in the inertial frame to compare the amplitude spectrum of KIC~11907454, one of the $\gamma$~Dor stars for which \citet{vanr16} found r~modes from their period spacings. The model adopts a rotation frequency of $1.35$~d$^{-1}$, consistent with $1.35\pm 0.02$~d$^{-1}$ obtained by \citet{vanr16}, and parameters consistent with the spectroscopic analysis by \citet{vanr15}. 

Predicted amplitude distribution curves of r~modes with $m=1$ and 2 are superposed on the amplitude spectrum in the bottom panel of Fig.\,\ref{fig:gamDor}. Solid and dashed lines are for even ($k=-2$) and odd ($k=-1)$ modes, respectively. \citet{vanr16} identified r~modes of this star from the period spacing distribution in the $1 - 1.2$~d$^{-1}$ range; these are even r~modes of $m=1$. 
Our model predicts odd r~modes of $(m,k) = (1,-1)$ to be most visible at $\sim\!\!0.5$~d$^{-1}$ and the Fourier spectrum of KIC 11907454 shows the presence of several peaks there. However, those frequencies (shown in black in Fig.\,\ref{fig:gamDor}) coincide with beat frequencies (i.e., $|\nu_i-\nu_j|$) between high-amplitude even r~modes of $(m,k)=(1,-2)$ around $\sim\!\!1.2$~d$^{-1}$ and g~modes of $m=-1$ around $\sim\!\!1.6 - 1.9$~d$^{-1}$.  This indicates that no visible odd r~modes are present in KIC 11907454.
We note that the amplitudes for even $m=2$ modes (around $2.5$\,d$^{-1}$) expected from energy equipartition to $m=1$ modes are much higher than the observed ones, although the expected frequency range agrees. This may indicate that less energy is given for $m=2$ r~modes, or that the effect of dissipation is stronger.
 
The predicted frequency range for $m=1$ even r~modes ($k=-2$) agrees well with the observed frequency group 
at $\sim$1.2~d$^{-1}$. Note that the frequency group at $\sim$1.7~d$^{-1}$ can be identified, from their period spacings, as prograde sectoral g~modes of  $(m,k)=(-1,0)$. 
\citet{vanr15} obtained $v\sin i=109.3 \pm 4.7$~km\,s$^{-1}$ for KIC 11907454, while the equatorial velocity of our model is $113.3$\,km~s$^{-1}$. This indicates an inclination angle of  $\sim$75$^\circ$. 

\subsubsection{KIC~7583663}

\begin{figure}
  \includegraphics[width=\columnwidth]{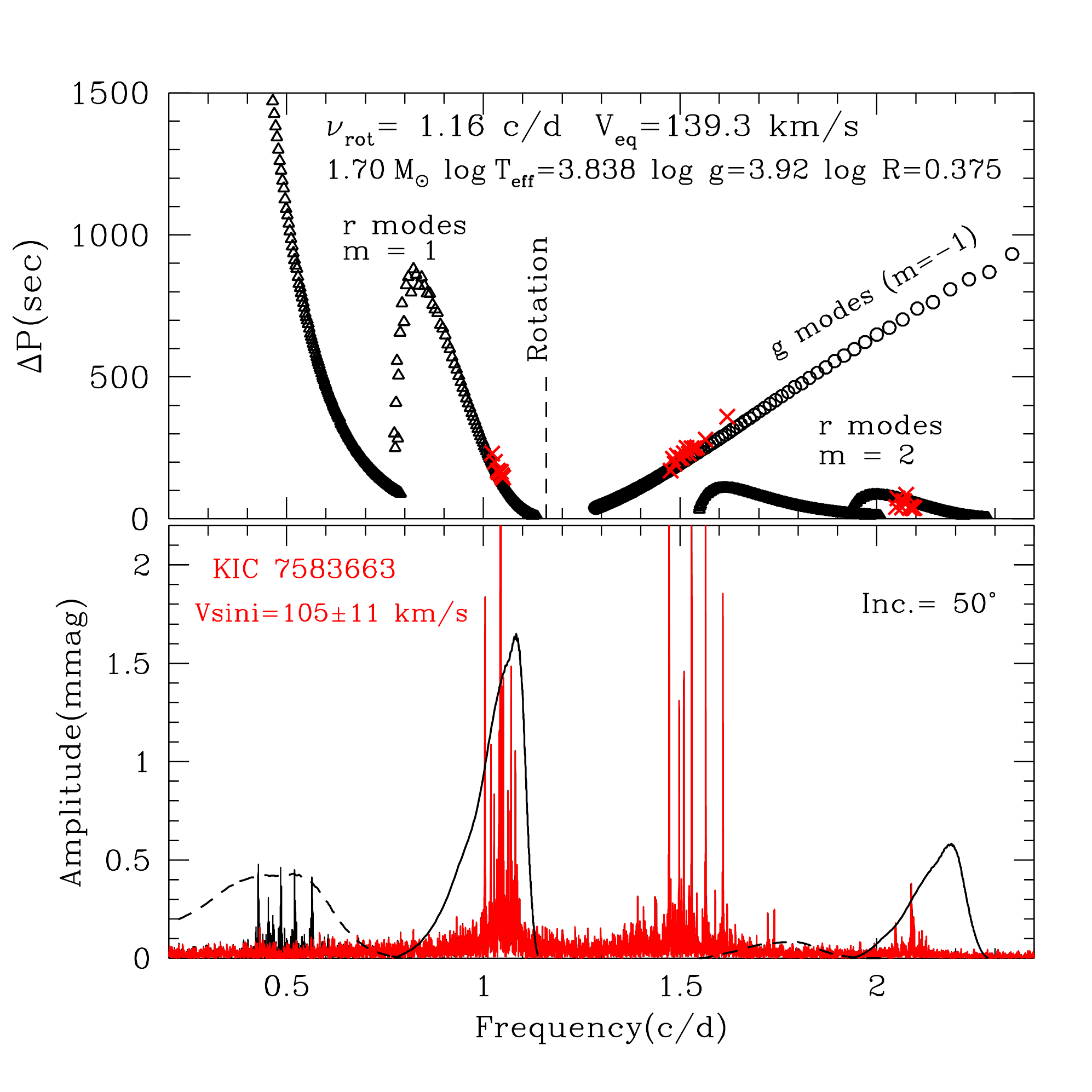}
\caption{The same as Fig.\,\ref{fig:gamDor} but for KIC~7583663, another $\gamma$~Dor star for which \citet{vanr16} found r~modes. Observational period spacings (crosses in the top panel) around 1 and 1.5~d$^{-1}$ are obtained by referring to fig.~33 of \citet{vanr15}, while period spacings in other parts are measured independently. \citet{vanr16} obtained a rotation frequency of    $1.17^{+0.02}_{-0.03}$\,d$^{-1}$ from period-spacings of r and g~modes. We have adopted $1.16$~d$^{-1}$ for the model which reproduces reasonably well (slightly better than a model with $1.17$~d$^{-1}$) the period spacings of r and g~modes, and amplitude distributions for even and odd modes of $m=1$ r~modes.  The inclination $50^\circ$ that was assumed is consistent with $v\sin i = 105\pm11$\,km\,s$^{-1}$ obtained by \citet{vanr15}.
The black part of the amplitude spectrum (bottom panel) indicates beat frequencies between r~modes of $(m,k)=(1,-2)$ and g~modes of $m=-1$.}
\label{fig:gamDor1.5}
\end{figure}

Fig.~\ref{fig:gamDor1.5} shows another example of a $\gamma$~Dor star, KIC~7583663, in which \citet{vanr16} found r~modes. We have adopted a rotation frequency of $1.16$~d$^{-1}$.  This is consistent with the value of $1.17^{
+0.02}_{-0.03}$~d$^{-1}$ preferred by \citet{vanr16}, but we found 1.16 d$^{-1}$ reproduces the observed g- and r-mode period spacings and amplitude distributions slightly better. The adopted rotation frequency and the radius of the model gives an equatorial rotation velocity of $139.3$\,km\,s$^{-1}$, while \citet{vanr15} obtained a spectroscopic velocity of $v\sin i = 105.0 \pm 11.0$\,km\,s$^{-1}$ for KIC~7583663. These numbers give an inclination angle of $i = 49^\circ\pm7^\circ$, which is somewhat lower than the previous case of KIC~11907454.

As in the previous case, \citet{vanr16} found the presence of r~modes from the period spacings around $\sim$1\,d$^{-1}$, which corresponds to the even ($k=-2$)  $m=1$ r~modes. 
Again, the frequency group around $\sim$0.5\,d$^{-1}$ is found to consist of beat frequencies between the even r~modes of $(m,k)=(1,-2)$ around $\sim$1\,d$^{-1}$, and g~modes of $m=-1$ around $\sim\!\!1.5$~d$^{-1}$, although the frequency range lies within the expected range of odd r~modes of $(m,k)=(1,-1)$. 
A small group around $2.1$\,d$^{-1}$ can be explained as even ($k=-2$) $m=2$ r~modes. 
In addition, a tiny group at $\sim$1.7\,d$^{-1}$ may consist of odd ($k=-1)$ $m=2$ r~modes, or it may be a part of the g~mode group. 

For the $\gamma$ Dor stars in which \citet{vanr16} found r~modes (the two stars discussed above and another eight stars in the Appendix), the predicted frequency ranges and amplitude distributions for even r~modes of $(m,k)=(1,-2)$ are found  to be consistent with the amplitude spectra of those stars. However, the amplitude distributions for $m=2$ even r~modes -- predicted from energy equipartition -- are found to be much larger than the observed frequency peaks in the corresponding frequency ranges for all cases. This may indicate that the energy is not partitioned equally between $m=1$ and $m=2$ modes.

The amplitude spectra of those stars (in particular KIC~11907454 and KIC~7583663) show frequency peaks in the range predicted for odd r~modes of $(m,k)=(1,-1)$.  However, those frequencies are always found to be identical to beat frequencies between even r~modes of $(m,k)=(1,-2)$ and g~modes of $m=-1$, or among the g~modes, indicating that no visible odd r~modes are present in those stars.

\subsubsection{KIC~5608334}

\begin{figure}
  \includegraphics[width=\columnwidth]{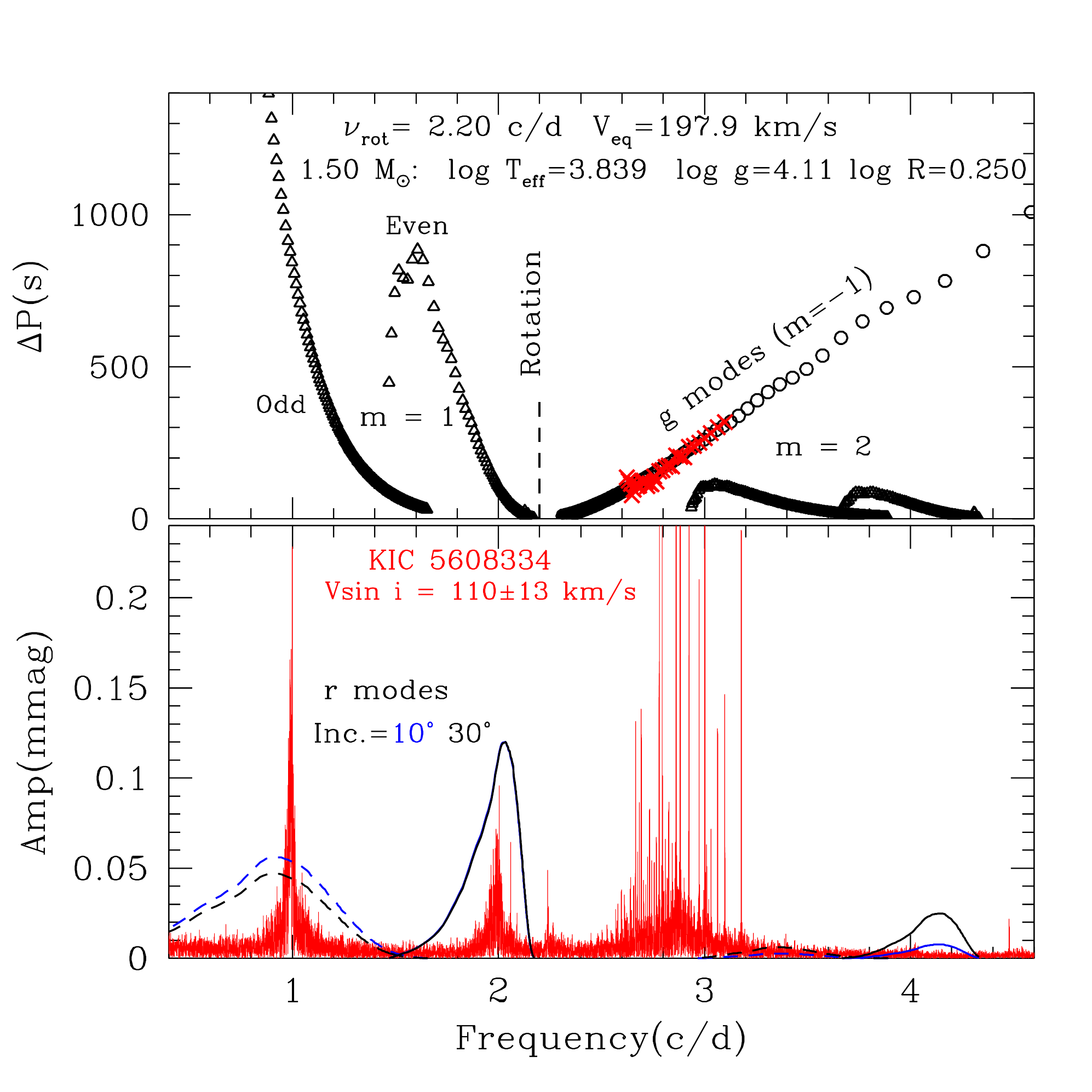}
\caption{The top panel shows predicted period spacings of r~modes (triangles) and prograde sectoral g~modes of $m=-1$ (circles) for a 1.5-$M_\odot$ model with a rotation frequency of $2.2$~d$^{-1}$ (indicated by a vertical dashed line). Period spacings of the $m=-1$ g~modes in the $\gamma$ Dor star KIC~5608334 are shown by crosses (period spacings of r~modes could not be measured because the frequencies are too densely distributed). 
The bottom panel compares the amplitude spectrum of  KIC~5608334  with the expected amplitude distributions of r~modes for assumed inclination angles of $10^\circ$ and $30^\circ$.  
The predicted amplitude distributions are normalised for the peak amplitude of $(m,k)=(1,-2)$ to be comparable to that of the hump at $\sim\!\!2$~d$^{-1}$; because of the normalisation, lines for the two inclination angles are nearly identical for $(m,k)=(1,-2)$. 
Note that only the low frequency part of the amplitude spectrum is shown; KIC~5608334 has additional extensive g-mode frequency groups above $5$~d$^{-1}$.
 }
\label{fig:gamDor2}
\end{figure}

Here we discuss another $\gamma$~Dor star, KIC~5608334; this star is not in the list of \citet{vanr16}. 
The top panel of Fig.\,\ref{fig:gamDor2}  shows predicted period spacings for r~modes (triangles),  and predicted (circles)  and measured (crosses) period spacings for prograde $m=-1$ g~modes. We have chosen a rotation frequency of $2.2$~d$^{-1}$ for the model by fitting the period spacings of the g~modes. Although a single peak at $2.24$~d$^{-1}$ and its harmonic (bottom panel of Fig.\,\ref{fig:gamDor2}) might be related to rotation, we found that  the g-mode period spacings of KIC~5608334 agree better with the rotation frequency $2.20$~d$^{-1}$ (rather than $2.24$~d$^{-1}$). 
The rotation frequency and the model radius predict an equatorial rotation velocity of $198$~km~s$^{-1}$, while \citet{nie15} obtained $v \sin i = 110 \pm 13$~km~s$^{-1}$ indicating an inclination of $34^\circ\pm5^\circ$.

The bottom panel  of  Fig.\,\ref{fig:gamDor2} compares a part of the amplitude spectrum of KIC~5608334  with model predictions of r~modes for inclinations of $30^\circ$ and $10^\circ$. 
Compared to the $\gamma$ Dor stars discussed above and in the Appendix, the pulsation amplitudes of KIC~5608334 are much smaller; the maximum amplitude of g~modes is about $0.6$~mmag, while the amplitudes of the previous 
$\gamma$~Dor stars are a few mmag. 
However,  the  peak  amplitude of KIC~5608334 at $\sim$1\,d$^{-1}$ corresponding to the odd ($k=-1$) r~mode group of $m=1$  is significantly higher than the peak amplitude at $\sim$2\,d$^{-1}$ corresponding to the even ($k=-2$) group. Such a ratio cannot be reproduced by our model, even with an inclination of as low as $\sim$10$^\circ$.
The $(m,k)=(1,-1)$ peak cannot be a group of beat frequencies between r~modes of $(m,k)=(1,-2)$ and $m=-1$ g~modes as in the $\gamma$ Dor stars discussed above, because frequency differences between the r~modes and the g~modes are mostly less than 1~d$^{-1}$, which is smaller than the frequency at the $(m,k)=(1,-1)$ peak.
The peak of the $1$-d$^{-1}$ hump is sharp compared to our model predictions. These features cannot be explained as long as energy equipartition is assumed as in our models, although the frequency range agrees with the prediction for odd ($k=-1$) r modes of $m=1$. One possible explanation might be the presence of some unknown resonance coupling to supply energy to the modes at $\sim$1\,d$^{-1}$.   

In contrast to the previous two $\gamma$~Dor stars, frequencies in the r-mode region of KIC~5608334 are too densely distributed to be resolved, even with the 4-yr {\it Kepler} data set. This is expected in many cases, as was explained in Section\,\ref{section:periodspacings}.

\subsection{Spotted A and B stars  -- `hump~\&~spike' stars} \label{sec:humpspike}

In amplitude spectra of Kepler light curves for mid-A to late-B stars, \citet{bal13,bal17} found a specific pattern 
(which he named the ROTD pattern, although no explanation of the acronym was given) of a broad spread of unresolved peaks and a sharper peak just to the high frequency side of the broad hump in many stars. He argued that the broad spread of peaks is caused by star spots with finite life-times in a latitudinally differentially rotating star and that the sharp peak might be caused by a reflection from a Jupiter-size orbiting body. He incorrectly characterised the ``sharp peak'' as being fully resolved with a sync function window pattern; for most stars there is, however, significant structure to this peak. That, and the fact that with the short periods involved many of the purported planets would transit (there are no transits seen in these stars), rules out this conjecture of planets commonly orbiting A stars with orbital periods of only days as the source of the ``sharp peak''. The interpretation of the broad spread of peaks as rotational requires strong differential rotation and abundant solar-like spots in stars with very thin surface convection zones. It also cannot account for other similar broad humps in the amplitude spectrum. There is also the $\gamma$~Dor star KIC~7661054, for which \citet{murphyetal2016a} show that a low-frequency broad spread of peaks cannot be attributed to rotation, a clear counter-example to the hypothesis of many spots and strong differential rotation as an explanation of the broad humps with sharper peaks on the high frequency side. 

Instead, we hypothesise that the sharper peak (which usually has structure and consists of some multiple peaks) is produced by a weak star spot or spots; thus this frequency represents the rotation frequency. In that, we follow the suggestion of \citet{bal13},  
but in this case only one or a few spots are required, they are weak since the photometric amplitudes are only 10s of $\umu$mag, and the frequency and/or amplitude modulation of this peak is slow relative to the 4-yr length of the {\it Kepler} data set, hence the relative sharpness of the peak. The requirement for this spot hypothesis in A and B stars is much less astrophysically demanding than the idea that the hump is the result of many spots and strong differential rotation. The magnetic A peculiar (Ap) stars have abundance spots that are stable over years and decades; these produce rotational light variations and have long been used to determine rotation periods for Ap stars. 
Evidence is also emerging that canonically non-magnetic B and A stars might have very weak magnetic fields that could support weak spots \citep[e.g.][]{blazere2016,nei17}.
The magnetic B and A stars have rotational light variations of hundredths of a magnitude; for the stars we are studying here the variations are only 10s of $\umu$mag. Hence the spot requirement is mild. 

More importantly, it is our interpretation that the broad hump is actually a group of frequencies of $m=1$ even ($k=-2$) r~modes. This naturally explains why the sharper peak always appears at the immediately higher frequency side of the broad peak; this is a fundamental relation between rotation frequency and the frequencies of $m=1$ even  ($k=-2$) r~modes. Since the name `ROTD' is based on a concept in contradiction to our explanation of the features, we use a phenomenological name, `hump~\&~spike' instead.

\subsubsection{KIC~622381} 

\begin{figure}
  \includegraphics[width=\columnwidth]{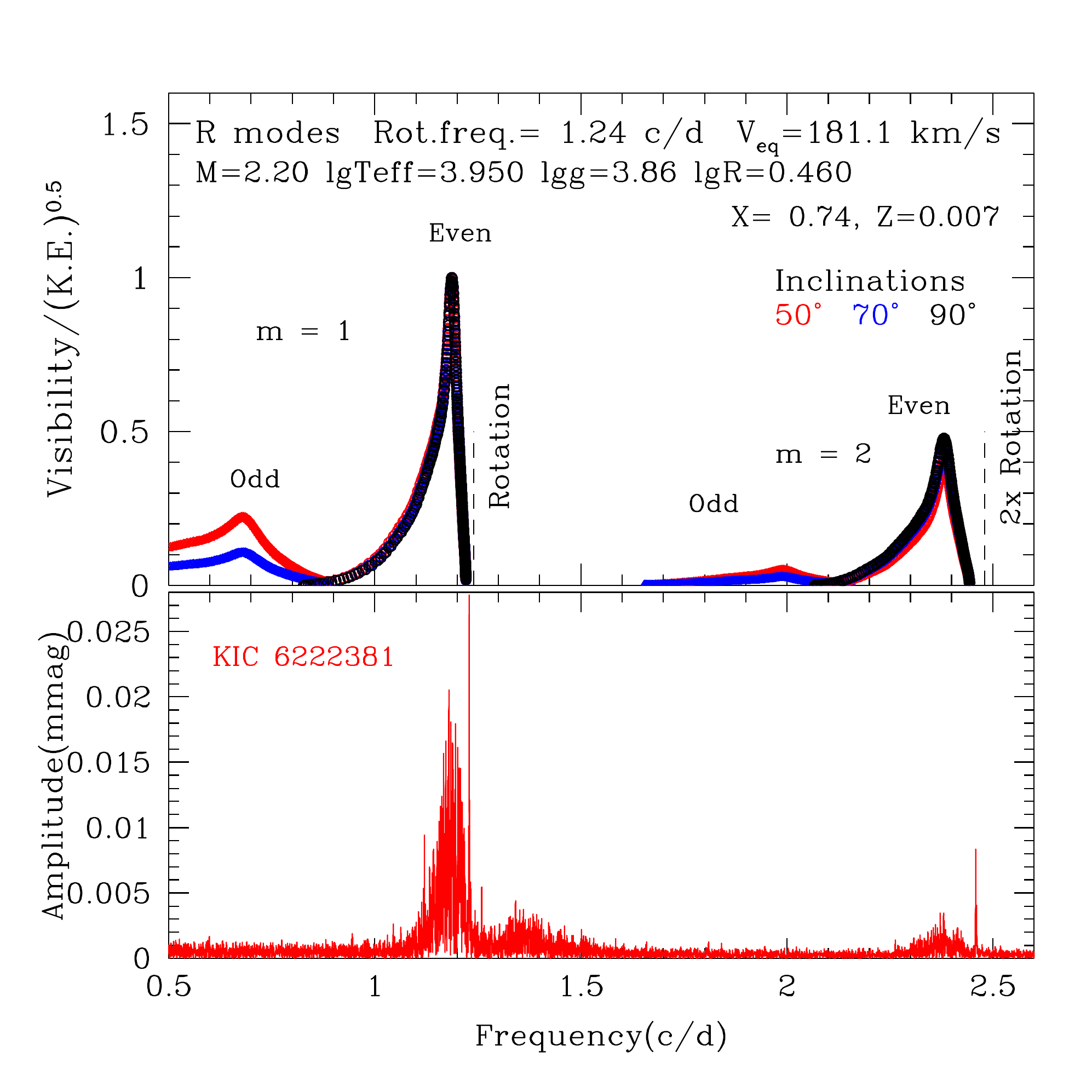}
\caption{ 
{\bf Top panel:} Visibility divided by the square root of the kinetic energy (Vis./$\sqrt{{\rm K.E.}}$; see eqs \ref{eq:vis} and \ref{eq:ke}) is plotted for three values of inclination ($50^\circ, 70^\circ, 90^\circ$) for a 2.2-M$_\odot$ model with a rotation frequency of $1.24$~d$^{-1}$. 
The visibility represents relative amplitudes expected for equipartition of kinetic energy among the modes. For each inclination the amplitude at the peak of even $m=1$ modes is normalized to unity. Although mode frequencies are plotted by open symbols (triangles and circles for odd and even modes, respectively), they are too densely distributed to be resolved. The normalized amplitude distribution of even modes barely depends on the inclination.
{\bf Bottom panel:} an amplitude spectrum of the {\it Kepler} 4-yr data set for KIC~6222381, one of the hump~\&~spike stars listed in \citet{bal17}, characterised by having a broad hump and a sharper peak at a slightly higher frequency. This star also has a harmonic of the main sharp peak, which shows some structure; we identify this sharp peak and its harmonic with the rotation frequency, and hypothesise that the light variations caused by a weak spot or spots on the surface.
The rotation frequency adopted in the top panel is nearly equal to the frequency of the main spike. We identify the broad humps around $\sim\!\!\!1.2$ and $\sim\!\!\!2.4$~d$^{-1}$ as even r modes of $m=1$ and $m=2$, respectively.  A low-amplitude hump around $\sim\!\!\!1.4$~d$^{-1}$ is probably a group of prograde dipole g modes.   
}
\label{fig:rotD}
\end{figure}

Fig.\,\ref{fig:rotD} shows an example of the hump~\&~spike star, KIC~6222381, a slightly evolved mid-A star with an effective temperature of $8700\pm300$~K and $\log g = 3.7$ \citep{hub14}. This star has a sharp peak at $1.229$~d$^{-1}$ \citep{bal17}, and a harmonic of this peak at twice that frequency. 
 We identify the sharp peak with the rotation frequency of the star with the light variations caused by a weak spot or spots on the surface. The amplitude of the sharp peak is only 28\,$\umu$mag; that of its harmonic is 8\,$\umu$mag. Hence the requirement is for a very weak spot, or spots. Prior to the {\it Kepler} data there are no observations of sufficient precision to have detected such small light variations. 

\begin{figure}
  \includegraphics[width=\columnwidth]{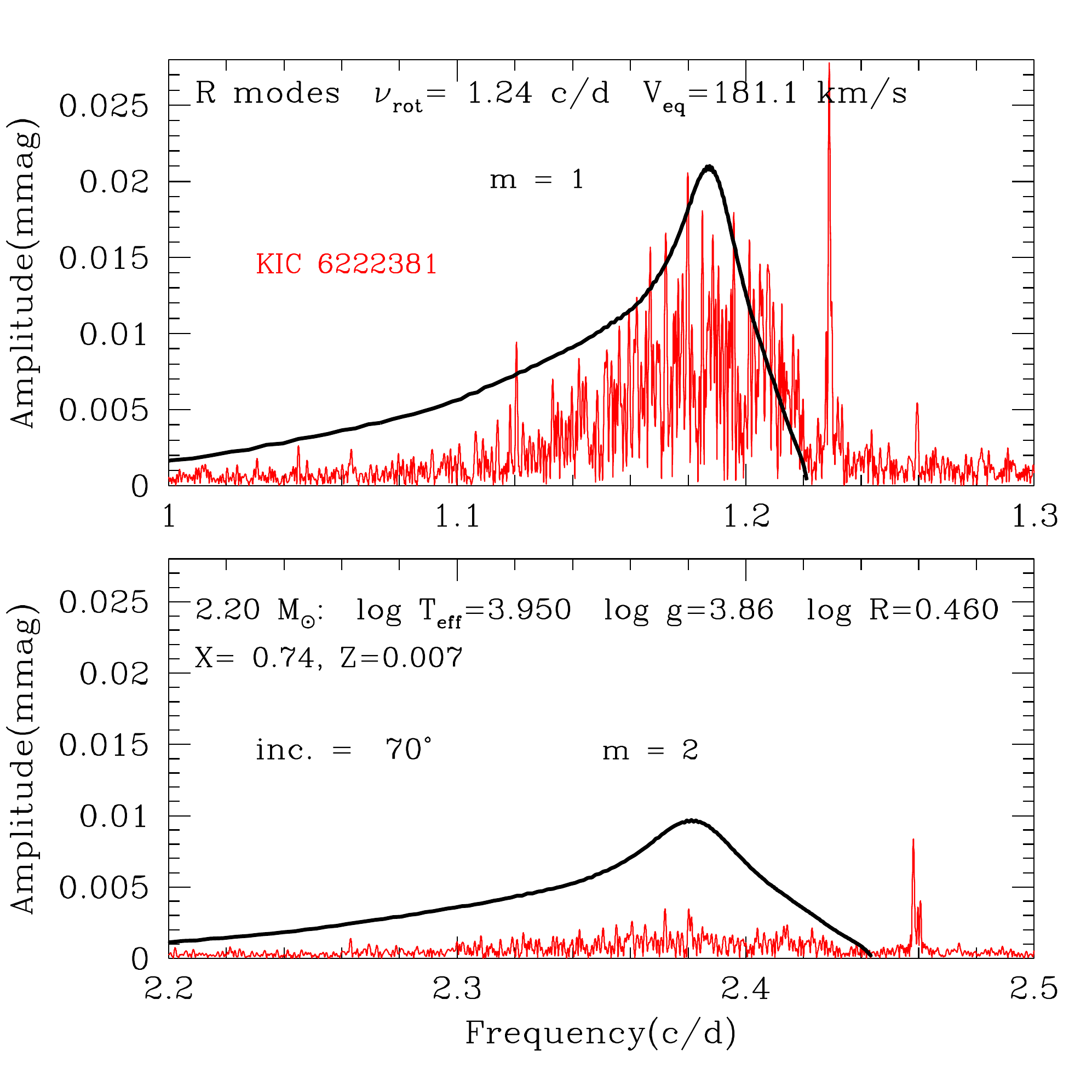}
\caption{Visibility distributions of even ($k=-2$) $m=1$ (top panel) and $m=2$ (bottom panel) r~modes are compared in detail with  the corresponding humps  in the amplitude spectrum of KIC 6222381. The model is the same as in Fig.~\ref{fig:rotD}, but frequency range of each group is zoomed. An inclination of $70^\circ$ was chosen here; the dependence of the visibility of even modes on the inclination is small as can be seen in Fig.~\ref{fig:rotD}.}
\label{fig:rotDzoom}
\end{figure}

To model the broad humps, we have adopted a slightly metal poor 2.2-M$_\odot$ model, taking into account $\rm{[Fe/H]}=-0.5 \pm 0.3$ as listed in \citet{hub14}, although metallicity hardly affects the results. Our model predicts that the $\sim\!\!1 - 1.2$~d$^{-1}$ group corresponds to $m=1$ even ($k=-2$) r~modes, while the group at $\sim$2.4\,d$^{-1}$ corresponds to $m=2$ even ($k=-2$) r~modes, reproducing well the broad distribution of peaks just to the low-frequency side of the main sharp peak.
Detailed comparison of each group is given in Fig.~\ref{fig:rotDzoom}; the expected frequency ranges agree well with the  features of both groups in the amplitude spectrum. 

The model presented, which best reproduces the observed frequency ranges of the two groups, has a rotation frequency of $1.24$~d$^{-1}$, which is very close to the frequency of the sharp peak, supporting the hypothesis that it is the rotation frequency. We note that in the zoomed diagrams (Fig.~\ref{fig:rotDzoom}) the sharp peak and its harmonic are broadened in low-amplitude parts indicating that the spot is not stable in this star over the 4-yr time span of the {\it Kepler} data set, and/or differential rotation may be present. 

In addition, there is a low-amplitude broad feature ranging from $1.3$ to $1.5$~d$^{-1}$ at the high-frequency side of the rotation frequency (see Fig.\,\ref{fig:rotD}). This frequency range relative to the rotation frequency suggests that the low-amplitude hump may be caused by a group of prograde sectoral g~modes of $m=-1$, in analogy of the $\gamma$~Dor stars KIC~11907454 in Fig.~\ref{fig:gamDor} and KIC~7583663 in Fig.~\ref{fig:gamDor1.5}. In contrast to those stars, however, the amplitudes of the possible g~modes of KIC~6222381 are much lower than those of the r~modes.

\subsubsection{KIC~6138789}

\begin{figure}
  \includegraphics[width=\columnwidth]{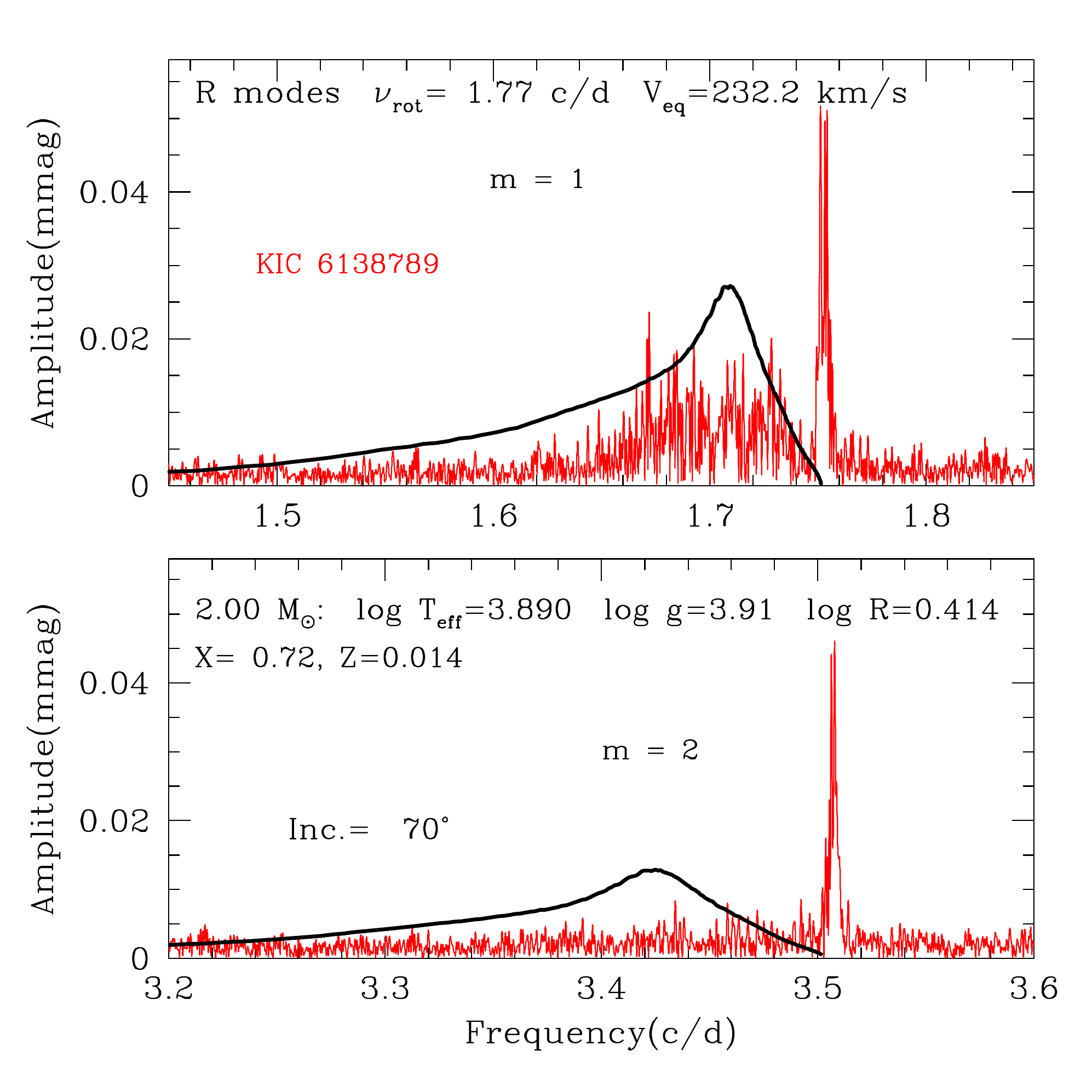}
\caption{Broad frequency groups with sharper peaks of KIC~6138789 are compared with predicted amplitude distributions for $m=1$ (top) and $m=2$ (bottom) even ($k=-2$) r~modes. The theoretical amplitude distribution is arbitrarily normalized.}
\label{fig:B}
\end{figure}

Fig.~\ref{fig:B} shows another A star, KIC~6138789, with sharp peaks and broad features. As in the previous case, the main broad feature around $\sim$1.7~d$^{-1}$ is explained by even r~modes of $m=1$ (top panel), and the small amplitude broad feature around $\sim$3.4~d$^{-1}$ as $m=2$ even r~modes (bottom panel). In contrast to the previous case, however, the sharp feature consists of two (or three) peaks, indicating the presence of two or three spots in different latitudes and a weak latitudinal differential rotation, or even just amplitude modulation caused by changes in the strength of a single spot. Irrespective of a single or a multiple sharp feature, the broad features are the same; the main group is identified as $m=1$ even r~modes, and the small amplitude group identified as $m=2$ even r~modes.

\subsubsection{KOI-81 (KIC~8823868)}

\begin{figure}
  \includegraphics[width=\columnwidth]{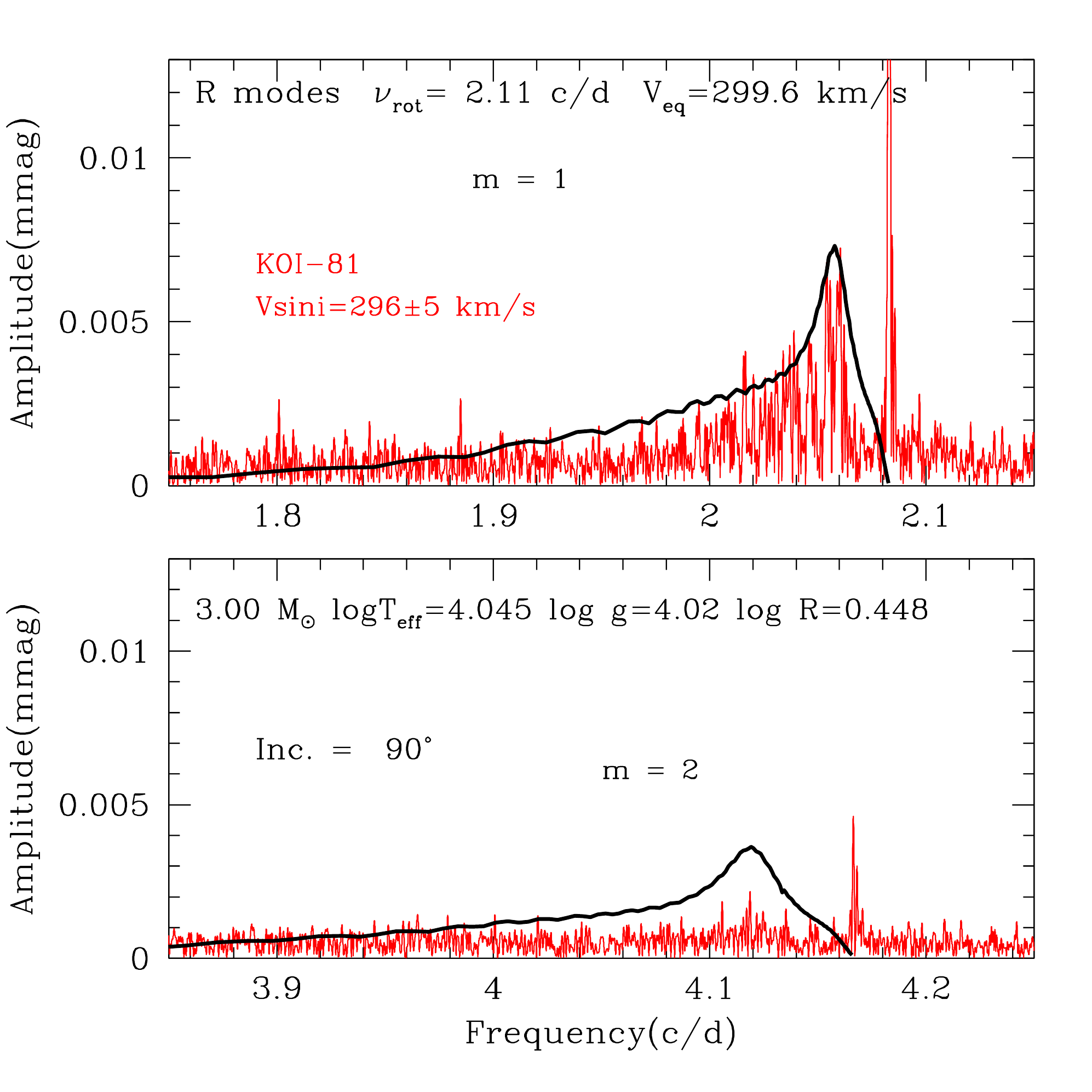}
\caption{The feature of a sharp peak and a broad group of peaks in the amplitude spectrum of the rapidly rotating B star KOI-81 (KIC~8823868) is compared with the visibility distribution of even ($k=-2$) r~modes for a 3.0-M$_\odot$  model, where the inclination is assumed to be  $90^\circ$. 
The top and the bottom panels are for the ranges of $m=1$ and $m=2$ r~modes, respectively. 
The theoretical visibility is normalized such that $m=1$ peak nearly agree with the broad hamp of KOI-81 around $\sim\!\!2.3$~d$^{-1}$. }
\label{fig:koi81}
\end{figure}

\citet{mat15} found a similar feature in the amplitude spectrum of the rapidly rotating B star KOI-81 (Fig.~\ref{fig:koi81}), which forms a 29.3-d eclipsing binary system with a subdwarf B star. They derived parameters of the primary B star to be: $M=2.92$\,M$_\odot$, $\log (R/{\rm R}_\odot) = 0.389$, $\log g = 4.13$, $\log {\rm T}_{\rm eff} = 4.068$, and $v\sin i = 296$~km~s$^{-1}$.

Fig.~\ref{fig:koi81} shows a comparison of the observed features with a model with parameters that are consistent with those obtained by \citet{mat15}. The adopted rotation frequency of $2.11$~d$^{-1}$ is slightly higher than the frequency at the sharp peak ($2.08$~d$^{-1}$). The equatorial velocity of the model, $300$~km~s$^{-1}$, indicates an inclination angle of $\sim$90$^\circ$, consistent with the observed eclipses and spin-orbit alignment.

It is known that the $\kappa$-mechanism at the Fe opacity bump excites intermediate order  r~modes in rotating SPB stars \citep{tow05,sav05,lee06}. For $m=1$ r~modes, \citet{tow05} and \citet{lee06} found only odd ($k=-1$) modes to be excited, while \citet{sav05} found a few even ($k=-2$) r~modes are also excited in a 3-M$_\odot$ model. We performed a non-adiabatic analysis for $m=1$ r~modes for the model shown in Fig.\,\ref{fig:koi81} using the method of \citet{lee95}, and found that many odd modes and a few even modes are excited, in agreement with the result of \citet{sav05}. 

However, these thermally excited modes cannot be responsible for the r~mode feature in KOI-81, because odd modes, even if present, cannot be seen at an inclination close to $90^\circ$, and the number of the excited even modes is too small.  Therefore, r~modes in KOI-81 should be caused by the same, but unknown, non-thermal mechanism as in the spotted A and B stars  (the hump~\&~spike stars).

\subsection{Frequently bursting Be stars} \label{sec:be}
Applying the hypothesis of mechanical generation of r~modes, we can further argue that r~modes may be present in rapidly rotating Be stars. Be stars intermittently undergo outbursts that eject mass. During an outburst, rotational flow on the surface would be disturbed, which might generate r~modes. For this reason r~modes might commonly be present in Be stars. 
In particular, some Be stars are known to have frequent minor outbursts, for which we expect to find signatures of r modes relatively easily.  We find that photometric results are published for two frequent outbursters; KIC~6954726 (StH$\alpha$~166) by \citet{bal15,riv16}, and HD~51452 (B0IVe) by \citet{nei12}; the latter was observed by the CoRoT satellite.

\subsubsection{KIC~6954726}
\begin{figure}
  \includegraphics[width=\columnwidth]{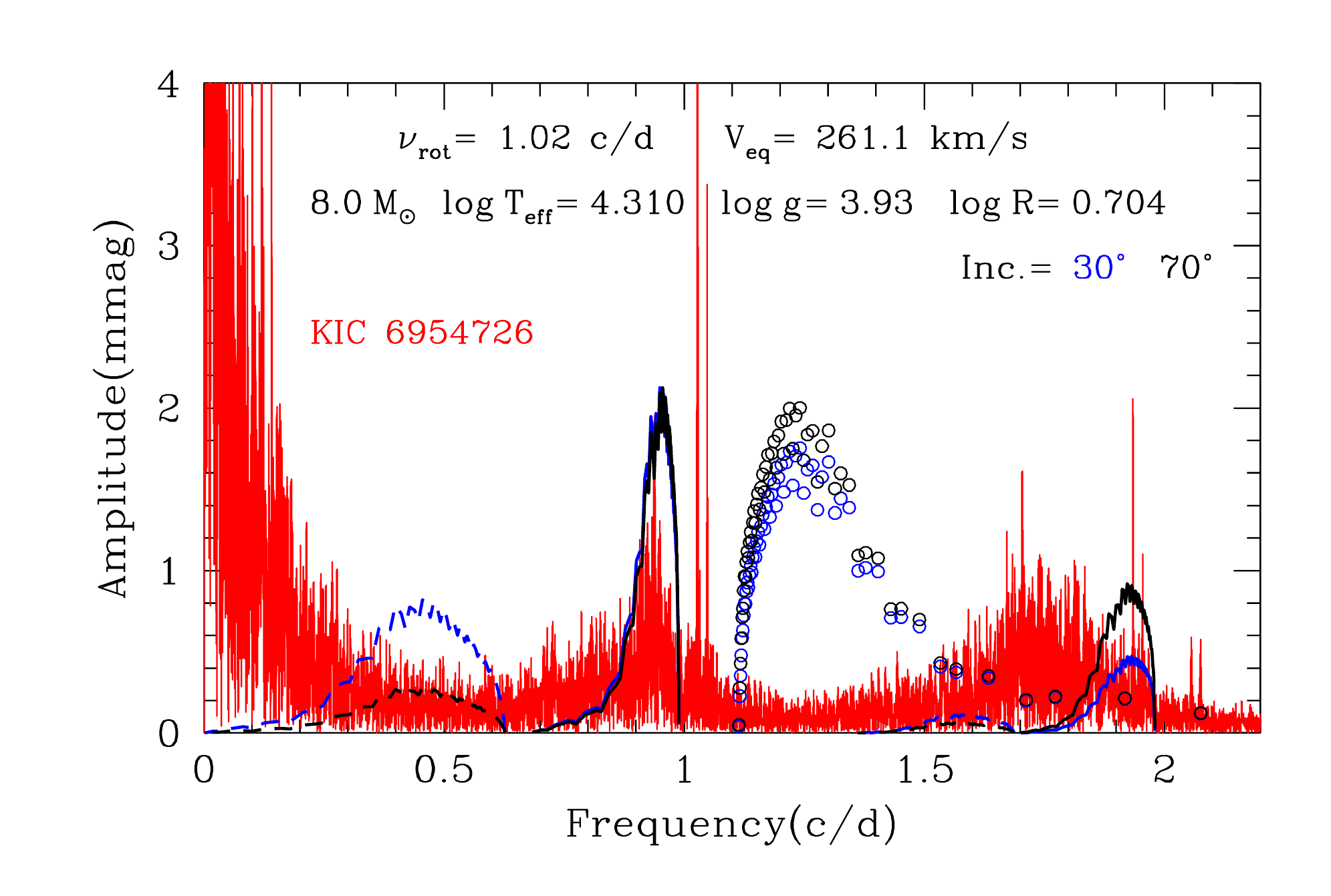}
\caption{A low-frequency part of  the amplitude spectrum of the Be star KIC~6954726 is compared with the visibilities of r~modes (solid ($k=-2$) and dashed ($k=-1$) lines) and of g~modes (open circles) for an 8-M$_\odot$ model with a rotation frequency of $1.02$~d$^{-1}$. The visibilities for each inclination are normalised at the peak of  the even r~modes of $(m,k)=(1,-2)$ at $\sim\!\!0.95$~d$^{-1}$. The low-frequency ($\la 0.2$~d$^{-1}$), high-amplitude features are caused by frequent outbursts of KIC~6954726.}
\label{fig:behump}
\end{figure} 

The effective temperature of KIC~6954726 is uncertain; \citet{riv16} estimated $17 < T_{\rm eff}({\rm kK}) < 25$ from metallic and He spectral lines,  while \citet{hub14} give $T_{\rm eff}\approx 17$~kK. \citet{bal15} obtained $v\sin i=160$~km~s$^{-1}$. Fig.~\ref{fig:behump} shows the amplitude spectrum  of KIC~6954726, which has a hump~\&~spike around $\sim\!\!1$~d$^{-1}$ and possibly a second one around $\sim\!\!2$~d$^{-1}$;
the high amplitude features in the low freqeuncy range $\la 0.2$~d$^{-1}$ are caused by frequent outbursts \citep[see][for light curves]{bal15,riv16}. The amplitudes of the humps are about $1-2$~mmag, much larger than those of typical hump~\&~spike stars (Sec.~\ref{sec:humpspike}), while comparable with the amplitudes of r modes in $\gamma$ Dor stars (Sec.~\ref{sec:gammaDor}).

Taking into account the $T_{\rm eff}$ range, and assuming the high amplitude spike at $\sim\!\!1$~d$^{-1}$ to be close to the rotation frequency of the star, we have calculated an 8-M$_\odot$ model at $T_{\rm eff} = 20.4$~kK with a rotation frequency of $1.02$~d$^{-1}$ for  KIC~6954726 . The predicted amplitude distributions are  compared with  the amplitude spectrum of KIC~6954726  in Fig.~\ref{fig:behump}. Solid and dashed lines are for even ($k=-2$) and  odd ($k=-1$) r~modes, respectively. For comparison, we show, by open circles, the visibilities of prograde sectoral dipole g modes ($(m,k)=(-1,0)$), which were calculated in the same way as for r~modes (eq.~\ref{eq:vis}).  This figure indicates that the hump around $0.9 - 1$~d$^{-1}$ agrees with the predicted frequency range of even r~modes of $(m,k)=(1,-2)$, and that the small hump at $\sim\!\!1.9$~d$^{-1}$ agrees with r~modes of $(m,k)=(2,-2)$. This strongly supports the presence of r~modes in the Be star KIC~6954726. However,  we do not know what caused the wide hump ranging $\sim\!\!1.6 - 1.85$~d$^{-1}$. Although it is located  within the g-mode frequency range of $(m,k)=(-1,0)$, the density of frequencies is much higher as seen in Fig.~\ref{fig:behump}.  \citet{riv16} argue that the broad hump, as well as another hump ranging $\sim\!\!0.7 - 1.1$~d$^{-1}$, are caused by aperiodic variations of a circumstellar disk.  Further discussion on the the origin of the broad humps is out of the scope of this paper, because various complex phenomena other than pulsations are occurring in a Be star.

\subsubsection{HD~51452}
Analysing CoRoT light curve for another frequent outburster, HD~51452,  \citet{nei12} obtained frequency groups around $\sim$0.5$\nu_{\rm rot}$ and $\sim$0.9$\nu_{\rm rot}$, where the rotation frequency  $\nu_{\rm rot}=1.22$~d$^{-1}$ was obtained spectroscopically. We find it likely that these frequencies are caused by $m=1$ odd and even r~modes, respectively. \citet{nei12}, on the othe hand, suggest that they are gravito-intertial modes including r modes which are excited stochastically in the convective core or in the envelope convection zone. One thing is clear though, these frequencies cannot be explained by thermally excited g~modes, because the effective temperature of a B0 type star is too high for the $\kappa$ mechanism at the Fe opacity bump to be effective for low-frequency modes.

\subsection{Heartbeat stars}
\begin{figure}
  \includegraphics[width=\columnwidth]{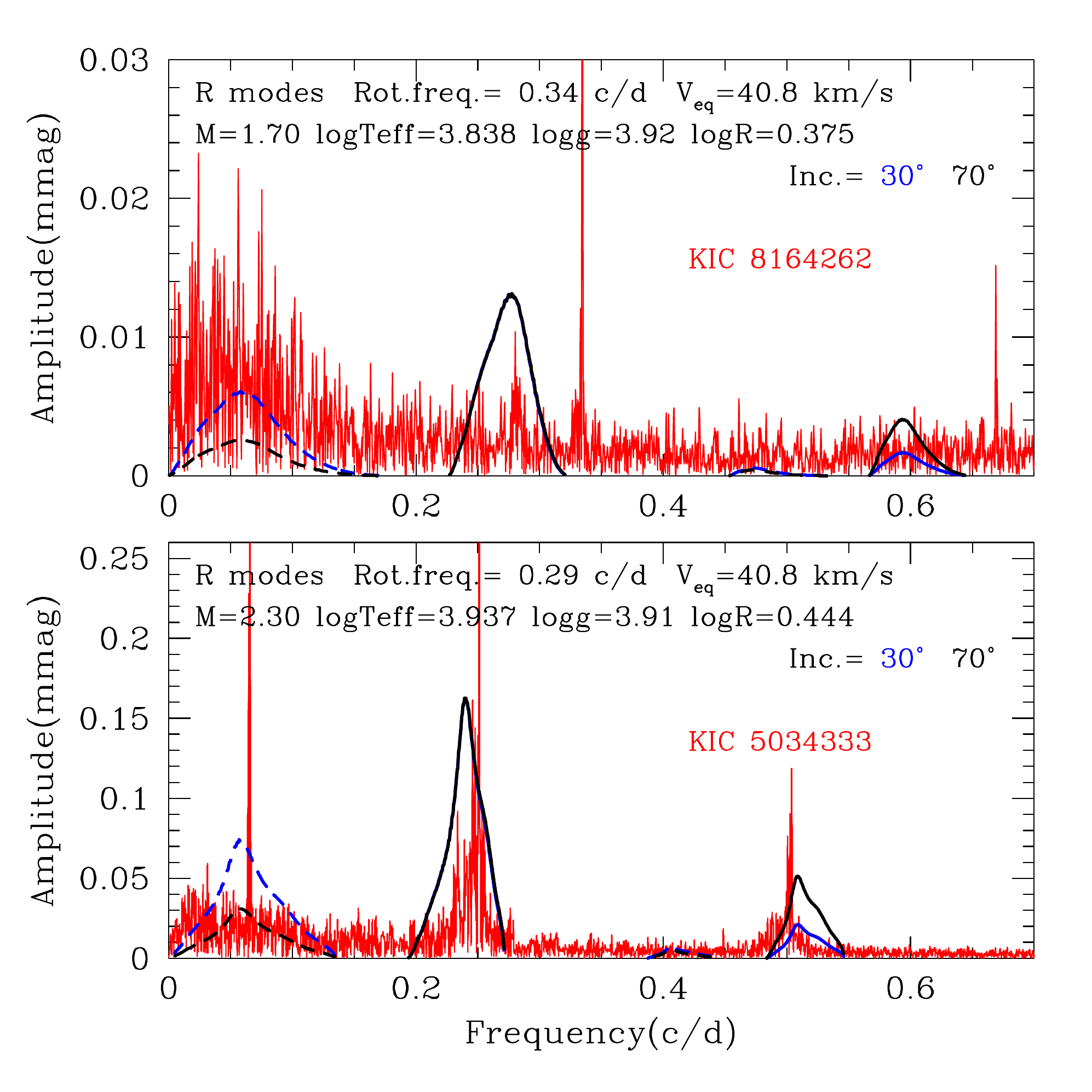}
\caption{Two examples of Heartbeat stars, KIC~8164262 (top) and KIC~5034333 (bottom). The rotation frequency adopted for KIC~8164262, $0.34$~d$^{-1}$ is very slightly higher than the frequency at the main single peak ($0.335$~d$^{-1}$). The rotation frequency for KIC~5034333 ($0.29$~d$^{-1}$) is estimated by fitting the visibility peak of $m=1$ even r~modes with the peak of the frequency group at $\sim$0.25~d$^{-1}$. Integer multiples of orbital frequencies have been removed from the amplitude spectrum.}
\label{fig:HB}
\end{figure}

Heartbeat stars are binary stars having highly eccentric orbits.  Strong tidal forces are exerted to  the outer structures of  the stars for a short span of time around each periastron passage. 
In this subsection, we discuss signatures in the amplitude spectra of some Heartbeat stars that indicate tidal disturbances may generate r~modes.

For the Heartbeat star KIC~8164262,  \citet{ham17} obtained three frequencies at $\sim$0.28~d$^{-1}$ (which can be recognised in a small frequency group in the amplitude spectrum in top panel of Fig.\,\ref{fig:HB}) not associated with rotation frequency nor with integer multiples of the orbital frequency. Although \citet{ham17} identified those frequencies as $\gamma$-Dor type g-mode oscillations, we consider that they are more likely  to be $m=1$ even ($k=-2$) r~modes (see below), which have frequencies just below the rotation frequency ($0.34$~d$^{-1}$) as in the cases of $\gamma$~Dor variables and spotted A and B stars. Furthermore, a hump in the low frequency range $\la 0.1$~d$^{-1}$ of KIC~8164262, if real\footnote{The potential for removal or alteration of astrophysical signals by the \textit{Kepler} data pipeline at frequencies less than 0.1\,d$^{-1}$ gives reason for caution. \citet{kur15} have shown that real signals below 0.1\,d$^{-1}$ are identifiable.}, might correspond to a group of $m=1$ asymmetric r~modes. Odd r~modes should be detectable, if excited, because a low inclination of $\sim$30$^\circ$ is estimated from $v\sin i=23\pm 1$~km~s$^{-1}$ \citep{ham17}. 

If the three frequencies found by \citet{ham17} at $\sim$0.28~d$^{-1}$ are g modes, they  may be excited thermally. (Tidal excitation is less likely for these frequencies because they are not equal to integer multiples of the orbital frequency.) The frequencies in the co-rotating frame should be $\sim\!\!0.62$~d$^{-1}$ if they are retrograde $m=1$ modes, and $\sim\!\!0.28$~d$^{-1}$ if they are zonal ($m=0$) modes. (They cannot be prograde modes because observed frequencies are less than the rotation frequency.)  Although the frequencies for the case of $m=1$ are within the range of g~modes excited thermally in $\gamma$ Dor stars \citep{dup05}, the thermal excitation of pulsation would be ineffective in a heartbeat star for the following reason. The thermal excitation works if the gas gains (loses) heat in the compressed (expanded) phase of pulsation \citep[e.g.][]{cox74}. The phase relation must be kept globally for a long time, because the growth time is much longer than the pulsation period itself. In the heartbeat stars, however, such a phase relation would be disrupted by strong tidal disturbances before an infinitesimal pulsation could grow to a visible amplitude. This supports our identification for the frequencies at $\sim$0.28~d$^{-1}$ to be r~modes of $(m,k)=(1,-2)$ generated by the tidal disturbances.

\citet{zim17} obtained rotation rates of many Heartbeat stars by assuming that the centre of relatively narrow features and their harmonics in the amplitude spectra are rotation frequencies and their harmonics. In most cases, however, the narrow feature is likely to be from a group of r~modes. An example, KIC~5034333, is shown in the bottom panel of Fig.\,\ref{fig:HB}. For KIC~5034333, \citet{zim17} give a rotation rate of $0.251$~d$^{-1}$ (i.e., a period of $3.98$~d) corresponding to the central peak of the main frequency group in the amplitude spectrum. On the other hand, identifying humps as groups of r~modes, we can fit the hump at $\sim$0.25~d$^{-1}$ with $m=1$ even ($k=-2$) r~modes of a model with a rotation frequency of $0.29$~d$^{-1}$, which is somewhat higher than the above value.   

Although the spike at $0.5$~d$^{-1}$ is likely the harmonic of the spike at $0.25$~d$^{-1}$, the small hump around the former spike is unlikely the harmonic of the main hump around the latter spike, because the amplitude distribution of each hump is different. In the model shown in the bottom panel of Fig.\,\ref{fig:HB}, the frequency range of the even $m=2$ r~modes nearly agrees with the small hump around $\sim$0.5~d$^{-1}$. However, we have to be cautious. The agreement might be accidental, because the frequency difference between the visibility peaks of $m=1$ and $m=2$ modes depends on stellar parameters, while humps with the 1:2 frequency ratio often appear in Heartbeat stars \citep{zim17}. Tidal forces, which are not considered in our models, might be responsible for producing such a nearly harmonic hump. 

Our~model identifies the small enhancement of amplitude between 0 and $\sim$0.1~d$^{-1}$, possibly including the sharp peaks at $0.07$~d$^{-1}$, as $m=1$ odd ($k=-1$) r~modes, while \citet{zim17} identified the latter feature as the rotation frequency of the secondary star of the binary system.  

Thus, we see r~modes even in relatively slowly rotating Heartbeat stars. In the excitation of r~modes, tidal forces must play an important role. 

\section{Discussion}

We found signatures of r~modes in the amplitude spectra of many $\gamma$~Dor stars, rapidly rotating spotted A and B stars, frequently bursting Be stars, and Heartbeat stars. The common feature of r~modes in the amplitude spectra is the presence of broad humps that appear immediately below the rotation frequency (or 2 times of the rotation frequency for $m=2$ modes). 
Such features agree well with the predictions of even $(k=-2)$ r~modes ($m=1$ modes in particular), although the amplitudes predicted  from energy equipartition for $m=2$ even modes tend to be higher than the  observed amplitudes. The amplitude problem in the $m=2$ modes seems to indicate a lack of equipartition and/or stronger dissipations in $m=2$ modes. 

Although odd r~modes are thermally excited more easily than even r~modes in SPB stars \citep{tow05,sav05,lee06}, we have not found clear evidence of {\it odd} ($k=-1)$ r~modes in our samples.
The r~modes \citet{vanr16} found in $\gamma$ Dor stars are identified as even r~modes of $(m,k)=(1,-2)$. Although the amplitude spectra of those stars have appreciable peaks in the range for the odd r~modes, most of them are identified as beat frequencies between the even r~modes and g~modes, or beat frequencies between the g~modes. The amplitude spectrum of another $\gamma$ Dor star, KIC~5608334, has a tight frequency group at the centre of the predicted frequency range for the odd r~modes of $(m,k)=(1,-1)$.  Although this is not a group of beat frequencies, the frequency group seems too sharp to be consistent with our theoretical prediction for a group of odd r~modes.   

\begin{figure}
  \includegraphics[width=0.49\columnwidth]{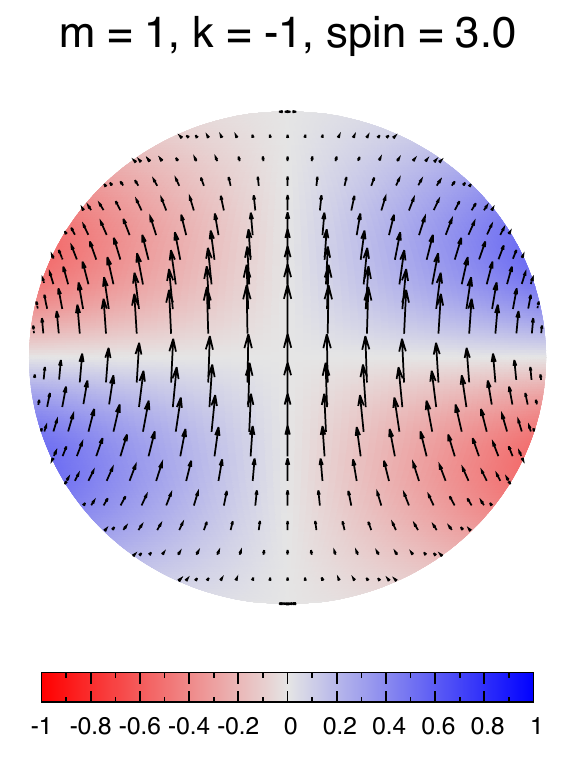}
  \includegraphics[width=0.49\columnwidth]{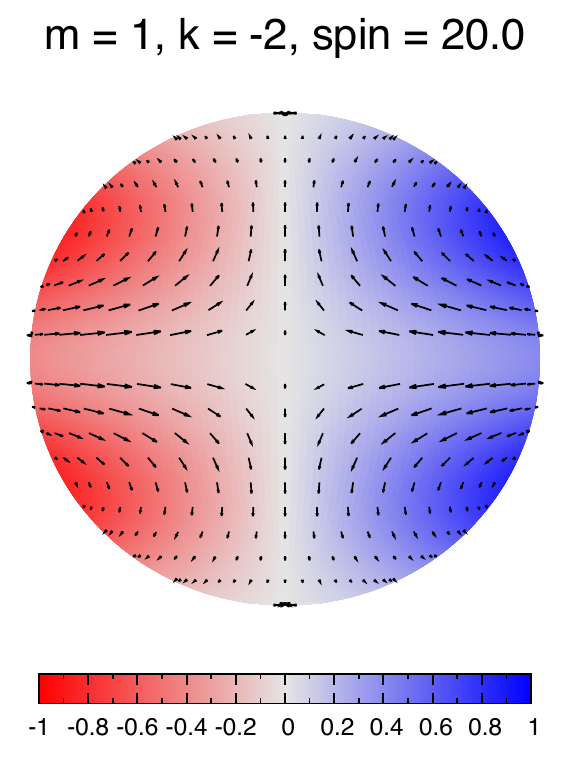}
 \caption{Flow patterns of an odd (left) and an even (right) r~mode of $m=1$ on the stellar surface at one pulsation phase are superposed on the temperature variations, where the inclination to the rotation axis is set to $90^\circ$ (the rotation axis points upward). Blue and red parts correspond to positive and negative temperature perturbations, respectively. 
Hotter (cooler) parts are associated with clockwise (counter clockwise) motions in the northern hemisphere, while  the correspondence is the opposite in the southern hemisphere. Parameters for each mode are given above the each pattern, where `spin' means the spin parameter defined as $2\nu_{\rm rot}/\nu^{\rm co}$. 
}
\label{fig:velocity}
\end{figure}

Our samples of spotted A and B stars (hump~\&~spike stars) have no frequency groups corresponding to odd r~modes. Furthermore,  no humps for the odd $m=1$ seem to be present in the samples of \citet{bal13,bal17}. 
These facts indicate that generating mechanically odd r~modes of $(m,k)=(1,-1)$ is difficult. The difficulty may come from the property of  the flow patters of  odd r~modes.  
As seen in Fig.\,\ref{fig:velocity}, the odd $m=1$ r~modes correspond to the largest scale motions involving a whole hemisphere crossing the equator, while the flows of even modes of $m=1$ are separated into the north and south hemisphere. 
 
If r~modes are excited mechanically by deviated flows from latitudinal differential rotation due to a blockage by a spot located at an intermediate to high latitude, we can imagine that $m=1$ even modes are most easily excited, while $m=1$ odd modes associated with flows crossing the equator are hardly excited. This is a reasonable explanation why no odd modes are detected in spotted A and B stars. 
In $\gamma$~Dor stars, g~modes might generate  $m=1$ odd (in addition to even) r~modes because the prograde sectoral g~modes have largest amplitudes on the equator. However, the amplitudes, even if excited, are expected to be small, because oscillation motions of the sectoral g~modes are symmetric to the equator (i.e. flows do not cross the equator), while the flows associated with the odd r~modes of $(m,k)=(1,-1)$ cross the equator (Fig.\,\ref{fig:velocity}).  

The hypothesis of mechanical generation of r~modes is supported by the presence of  r-mode humps in the frequently outbursting Be star KIC~6954724 (\S\ref{sec:be}) and possibly in HD~51452; the latter was observed by the {\it CoRoT} satellite. 

It is reasonable to imagine that r~modes in Heartbeat stars are excited by asynchronous tidal forces. In Heartbeat stars, we find $m=1$ r~modes and possibly small amplitude even $m=2$ r~modes whose frequencies might be affected by tidal forces. 

\section{Conclusions}

We modelled amplitude distributions of r~modes as a function of frequency, assuming equipartition of energy among r~modes. We found that the predicted amplitude distributions are consistent with the frequency ranges of even r~modes for $\gamma$~Dor stars. Applying such models to various stars, we found signatures of r~modes in the amplitude spectra in many cases of $\gamma$~Dor stars, spotted A and B stars,   and Heartbeat stars. We also found such a signature in one Be star observed by the {\it Kepler} satellite.  Our findings explain a broad hump associated with a sharper feature at slightly higher frequency that appears in the amplitude spectra of many early type stars. We have argued that these r~modes are generated mechanically by g-mode oscillations, by deviated flows caused by spotsor mass outbursts, and by non-synchronous tidal forces. We suggest that r~modes are potentially ubiquitous in moderately to rapidly rotating stars and may play an important role in transporting angular momentum in stars.

\section*{Acknowledgements}
We thank professor Hiromoto Shibahashi for useful discussions and comments on a draft of the paper. 
We thank professor Conny Aerts and the anonymous referee for useful comments.
We also thank NASA and the Kepler team for their revolutionary data.
SJM is supported by the Australian Research Council.
Funding for the Stellar Astrophysics Centre is provided by The Danish National Research Foundation (Grant agreement no.: DNRF106)




\bibliographystyle{mnras}
\bibliography{ref} 



\appendix

\section{Additional $\gamma$~Dor stars with r~modes}
\begin{figure*}
  \includegraphics[width=\columnwidth]{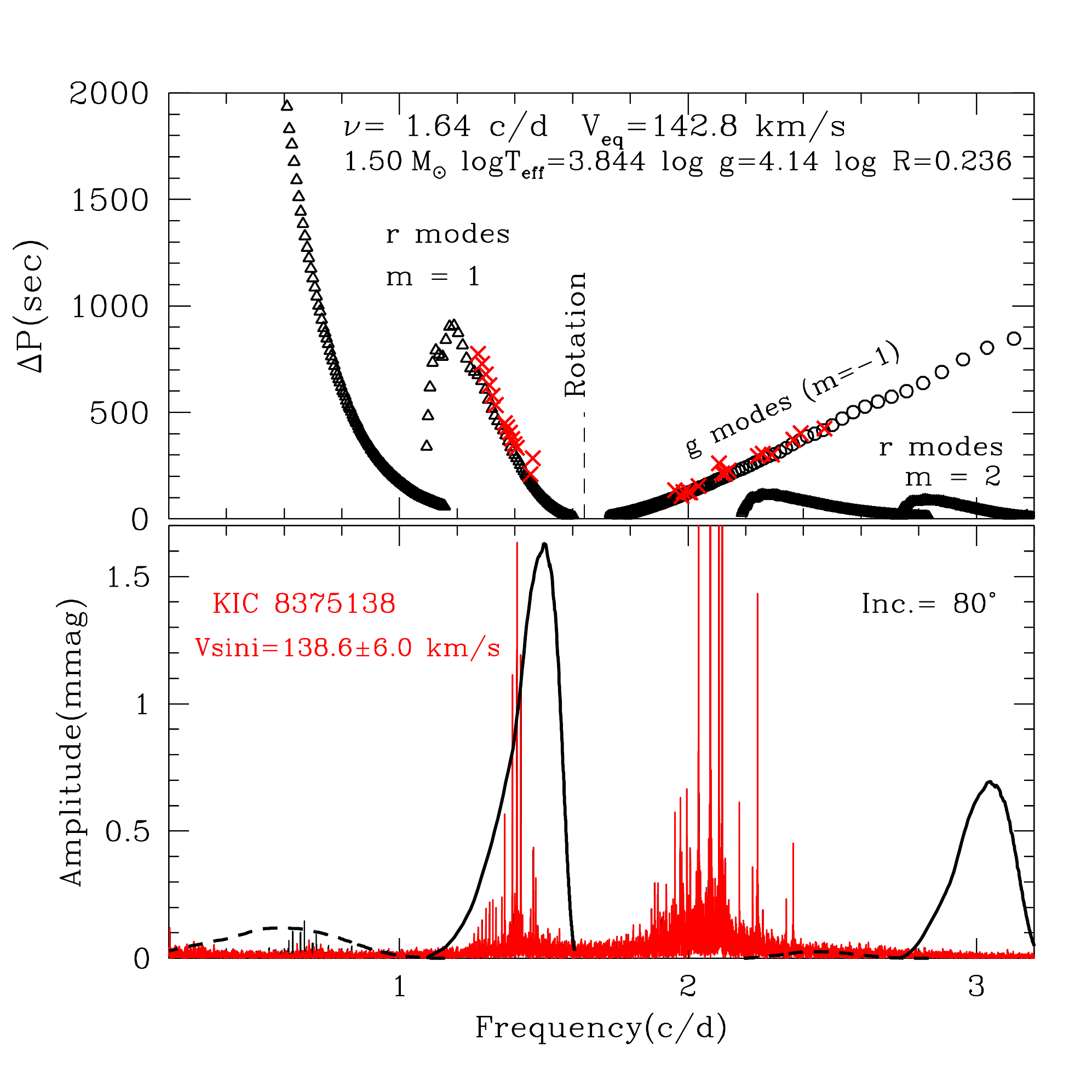}
  \includegraphics[width=\columnwidth]{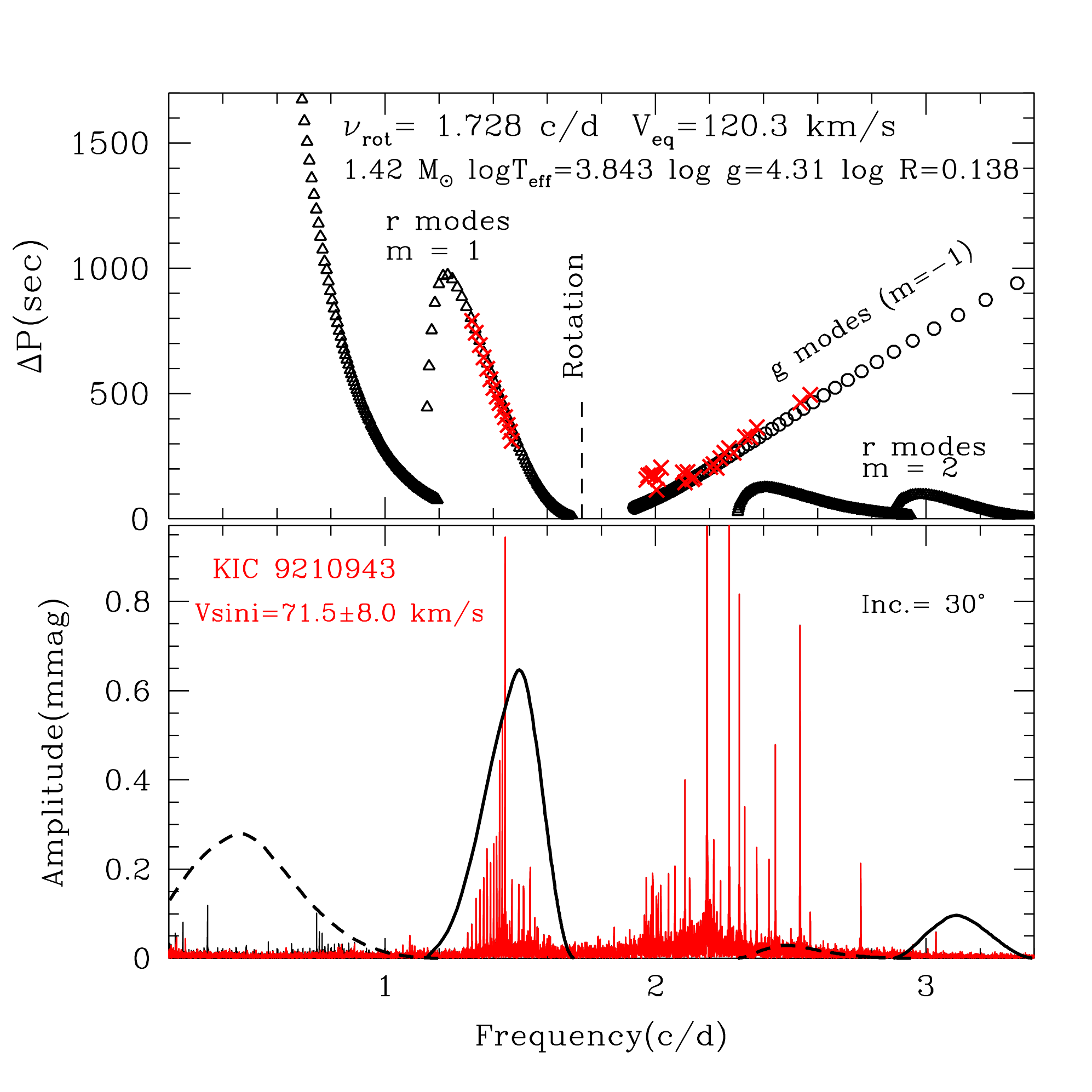}
\caption{Two examples of $\gamma$~Dor stars, KIC~8375138 (left) and KIC~9210493 (right), which show clean frequency spacinges of r~modes around $\sim$1.3\,d$^{-1}$. (The format of the diagrams is the same as Fig.\,\ref{fig:gamDor}.) These frequency (or period) spacings nicely agree with those of even ($k=-2$)  $m=1$ r~modes in models with rotation frequencies consistent with those obtained by \citet{vanr16}: $1.64^{+0.01}_{-0.01}$\,d$^{-1}$ for KIC\,8375138 and $1.728^{+0.007}_{-0.010}$\,d$^{-1}$ for KIC\,9210943. Values of $v \sin i$  are adopted from \citet{vanr15}. Small frequency peaks shown in black, in the frequency ranges expected for odd r~modes of $(m,k)=(1,-1)$,  are beat frequencies formed between even r~modes of $(m,k)=(1,-2)$ and g~modes of $m=-1$, or  among the high amplitude g~modes themselves.}
\label{fig:2gammDor}
\end{figure*}

\begin{figure*}
  \includegraphics[width=\columnwidth]{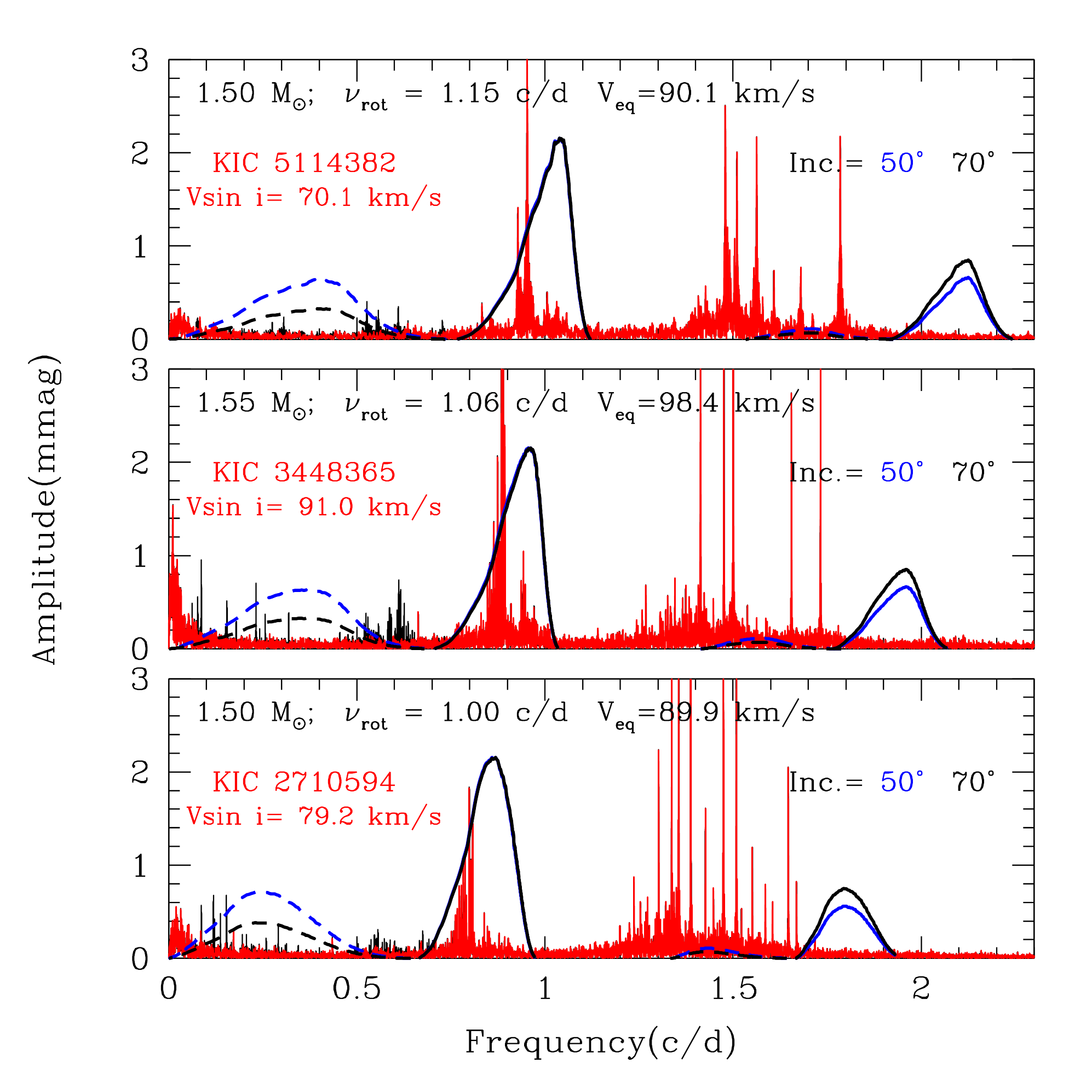}
  \includegraphics[width=\columnwidth]{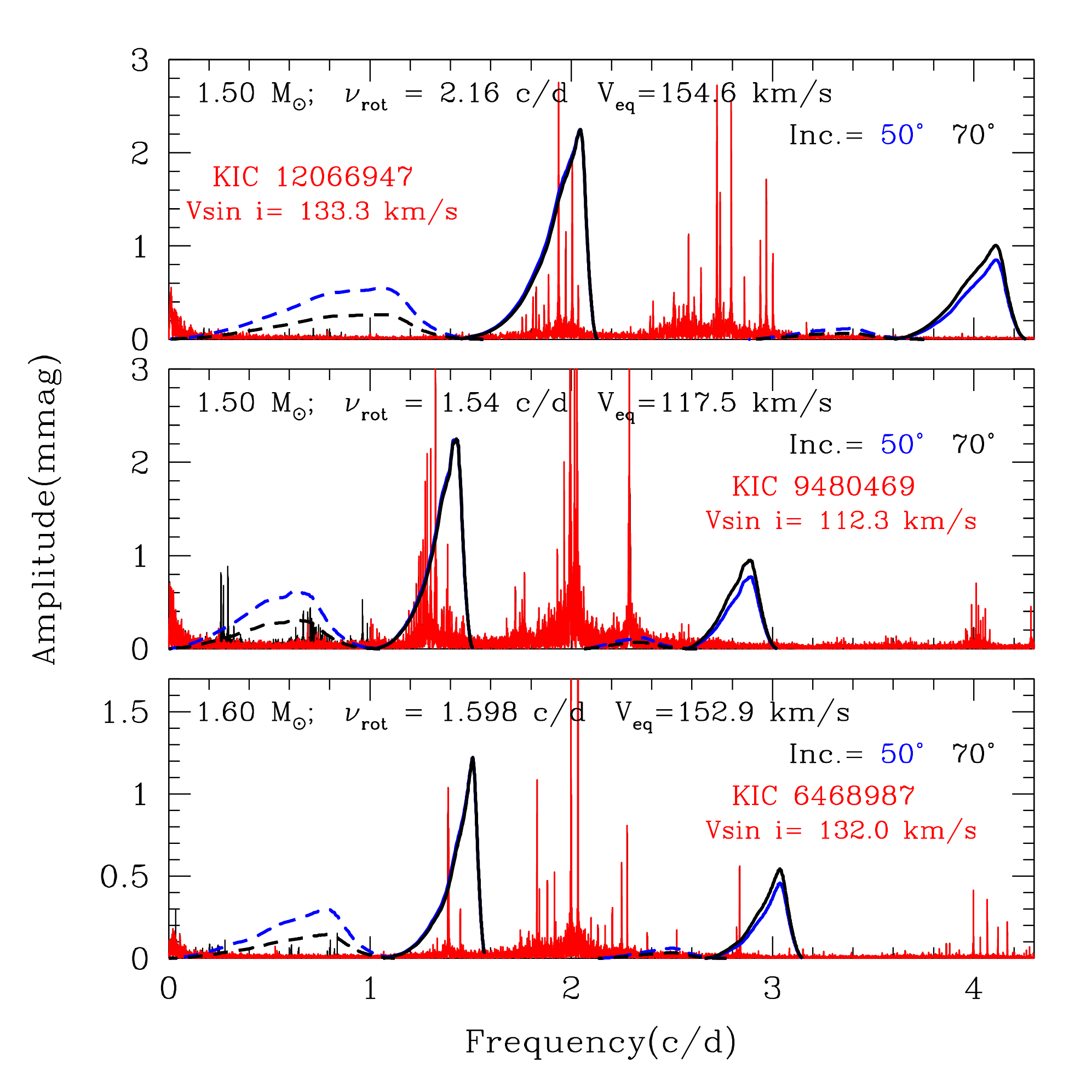}
\caption{Amplitude-frequency diagrams of six $\gamma$~Dor stars with r~modes are compared with model predictions for amplitude distributions of r~modes at inclinations of $50^\circ$ (blue lines) and $70^\circ$ (black lines), where solid and dashed lines are even ($k=-2$) and odd ($k=-1$) modes, respectively. The assumed rotation frequency, $\nu_{\rm rot}$, of each star is consistent with the result of \citet{vanr16}.  The frequencies of r~modes with azimuthal order $m$ are confined into the range between $(m-1)\nu_{\rm rot}$ and $m\nu_{\rm rot}$.
Even $m=1$ modes have maximum amplitudes among r~modes, and the amplitude is normalized arbitrarily at the maximum. 
(The blue and black lines for the even $m=1$ modes run almost identically.)
Frequency peaks shown in black are beat frequencies formed between even r~modes of $(m,k)=(1,-2)$ and g~modes of $m=-1$, or  among the high amplitude g~modes themselves.}
\label{fig:6gammDor}
\end{figure*}

\citet{vanr16} found r~modes in ten $\gamma$~Dor stars by looking at period spacing versus period relations.
We have already discussed two of the stars in \S\ref{sec:gammaDor}. In this Appendix we show model fittings for the remaining eight stars. Among them, KIC~8375138 and KIC~9210943 show clean frequency (period) spacings, which are nicely fitted with even ($k=-2$) m=1 r~modes as shown in Fig\,\ref{fig:2gammDor} with model predictions of amplitude distribution as a function of frequency. For the other six stars, the amplitude-frequency diagrams are compared with model predictions for inclinations of $50^\circ$ and $70^\circ$. In all cases, models adopt parameters consistent with those obtained by \citet{vanr15} and rotation frequencies obtained by \citet{vanr16}. Agreements between observational amplitude-frequency diagrams and model predictions for even ($k=-2$) r~modes are satisfactory for all cases except for that the amplitudes of $m=2$ even r~modes predicted with energy equipartition are higher than the observational amplitudes. 

Although several small amplitude peaks are seen in the ranges predicted for $m=1$ odd ($k=-1$) r~modes, most of them can be identified as beat frequencies between even r~modes of $(m,k)=(1,-2)$ and g~modes of $m=-1$, or beat frequencies among high amplitude g~mode frequencies. This indicates  that no visible odd r~modes are present in those $\gamma$ Dor stars.  
 

\bsp	
\label{lastpage}
\end{document}